\documentclass[prd,eqsecnum,twocolumn,amsfonts,amssymb]{revtex4}

\usepackage{CJK}

\usepackage{graphicx}

\usepackage{bm}

\newcommand{\beq}{\begin{equation}}
\newcommand{\eeq}{\end{equation}}
\newcommand{\beqs}{\begin{eqnarray}}
\newcommand{\eeqs}{\end{eqnarray}}
\newcommand{\lsim}{\mathrel{\raisebox{-
.6ex}{$\stackrel{\textstyle<}{\sim}$}}}

\newcommand{\drawsquare}[2]{\hbox{%
\rule{#2pt}{#1pt}\hskip-#2pt
\rule{#1pt}{#2pt}\hskip-#1pt
\rule[#1pt]{#1pt}{#2pt}}\rule[#1pt]{#2pt}{#2pt}\hskip-#2pt
\rule{#2pt}{#1pt}}
\newcommand{\fund}{\raisebox{-.5pt}{\drawsquare{6.5}{0.4}}}
\newcommand{\sym}{\raisebox{-.5pt}{\drawsquare{6.5}{0.4}}\hskip-0.4pt%
        \raisebox{-.5pt}{\drawsquare{6.5}{0.4}}}
\newcommand{\asym}{\raisebox{-3.5pt}{\drawsquare{6.5}{0.4}}\hskip-6.9pt%
        \raisebox{3pt}{\drawsquare{6.5}{0.4}}}

\begin{document}

\begin{CJK*}{UTF8}{}

\title{$A_k \, \bar F$ Chiral Gauge Theories}

\author{Yan-Liang Shi (\CJKfamily{bsmi}石炎亮) and Robert Shrock} 

\affiliation{C. N. Yang Institute for Theoretical Physics, 
Stony Brook University, Stony Brook, N. Y. 11794 }

\begin{abstract}

We study asymptotically free chiral gauge theories with an SU($N$) gauge group
and chiral fermions transforming according to the antisymmetric rank-$k$ tensor
representation, $A_k \equiv [k]_N$, and the requisite number, $n_{\bar F}$, of
copies of fermions in the conjugate fundamental representation, $\bar F \equiv
\overline{[1]}_N$, to render the theories anomaly-free. We denote these as $A_k
\, \bar F$ theories. We take $N \ge 2k+1$ so that $n_{\bar F} \ge 1$.  The $A_2
\, \bar F$ theories form an infinite family with $N \ge 5$, but we show that
the $A_3 \, \bar F$ and $A_4 \,\bar F$ theories are only asymptotically free
for $N$ in the respective ranges $7 \le N \le 17$ and $9 \le N \le 11$, and
that there are no asymptotically free $A_k \, \bar F$ theories with $k \ge 5$.
We investigate the types of ultraviolet to infrared evolution for these $A_k \,
\bar F$ theories and find that, depending on $k$ and $N$, they may lead to a
non-Abelian Coulomb phase, or may involve confinement with massless
gauge-singlet composite fermions, bilinear fermion condensation with dynamical
gauge and global symmetry breaking, or formation of multifermion condensates
that preserve the gauge symmetry. We also show that there are no asymptotically
free, anomaly-free SU($N$) $S_k \, \bar F$ chiral gauge theories with $k \ge
3$, where $S_k$ denotes the rank-$k$ symmetric representation.

\end{abstract}


\maketitle

\end{CJK*}


\section{Introduction}
\label{intro}

The properties of chiral gauge theories, especially in the strong-coupling
regime, remain a challenge for theoretical understanding.  One requires that
such a theory must be free of any triangle anomaly in gauged currents, since
such an anomaly would spoil the renormalizability of the theory.  Imposing the
additional requirement that such a theory must be asymptotically free
guarantees that there is at least one region, namely the deep ultraviolet (UV)
at large Euclidean energy/momentum scales $\mu$, where the running gauge
coupling $g(\mu)$ is small, so that the properties of the theory are reliably
calculable using perturbative methods.  A chiral gauge theory is said to be
irreducibly chiral if it does not contain any vectorlike subsector.  In this
case, the chiral gauge symmetry precludes any fermion mass terms in the
underlying Lagrangian.  We shall focus on irreducibly chiral theories here. In
an asymptotically free gauge theory, as the reference scale $\mu$ decreases
toward the infrared (IR) from the ultraviolet, the running gauge coupling
grows.  One possibility is that the beta function has an infrared zero at a
small value of the coupling, which constitutes an infrared fixed point
of the renormalization group (RG).  In this case, in the full
renormalization-group evolution of the theory from the deep UV to the IR, the
gauge interaction remains weakly coupled, and one expects that the IR behavior
is that of a (deconfined) non-Abelian Coulomb phase. In contrast, if the beta
function has an IR zero at a sufficiently large value of the coupling, or if
the beta function does not have any IR zero, then the theory becomes strongly
coupled in the infrared.  In this case, several types of behavior can occur.
Since the fermions are massless, the theory is invariant under a global flavor
symmetry. If one can construct gauge-singlet fermionic operator products that
match the global anomalies of the fundamental fermions, a condition known as 't
Hooft global anomaly matching, then the strongly coupled chiral gauge
interaction may confine and produce massless gauge-singlet composite spin-1/2
fermions \cite{thooft79}-\cite{cgt2}.  Alternatively, the strong gauge
interaction may produce fermion condensate(s) that spontaneously break gauge
and global chiral symmetries \cite{georgi86,eppz},
\cite{weinberg76}-\cite{cgtsab}. This latter type of behavior can occur in
several stages at different energy scales, resulting in a hierarchy of
symmetry-breaking scales.  In addition to their intrinsic field-theoretic
interest, strongly coupled chiral gauge theories have been applied in efforts
to construct (preon) models of composite quarks and leptons and models
explaining electroweak symmetry breaking and the structure of fermion
generations and masses in the Standard Model.  Some work along these lines
includes \cite{thooft79}-\cite{net}.

In this paper we shall study asymptotically free chiral gauge theories in four
spacetime dimensions (at zero temperature) with an SU($N$) gauge group and
chiral fermions transforming according to the antisymmetric rank-$k$ tensor
representation, denoted $A_k \equiv [k]_N$, and the requisite number, 
$n_{\bar F}$, of copies of fermions in the conjugate fundamental 
representation, $\bar F \equiv \overline{[1]}_N$, to render the theories 
anomaly-free \cite{conjugates}.  We denote these as $A_k \, \bar F$ theories. 
We take $N \ge 2k+1$ so that the theory is chiral and $n_{\bar F} \ge 1$.  
We extend previous studies on the $A_2 \, \bar F$ theories
\cite{drs,georgi86,eppz,dfcgt,ads,cgt,uvcomplete_etc} with further analysis of
fermion condensation channels and sequential symmetry breaking, while the $A_k
\, \bar F$ theories with $k \ge 3$ are, to our knowledge, new here.  The $A_2
\, \bar F$ theories form an infinite family with $N \ge 5$, but we show that
the $A_3 \, \bar F$ and $A_4 \,\bar F$ theories are only asymptotically free
for $N$ in the respective ranges $7 \le N \le 17$ and $9 \le N \le 11$, and
there are no asymptotically free $A_k \, \bar F$ theories with $k \ge 5$.  We
investigate the types of ultraviolet to infrared evolution for these $A_k \,
\bar F$ theories and find that, depending on $k$ and $N$, they may lead to a
non-Abelian Coulomb phase, or may involve confinement with massless
gauge-singlet composite fermions, or fermion condensation with dynamical gauge
and global symmetry breaking.  One of the methods that we use for our analysis
is the most-attractive-channel criterion for bilinear fermion condensation
\cite{drs}. We find two cases in each of which two channels are equally
attractive, so the most-attractive channel criterion cannot determine which is
more likely to occur; for these we show how vacuum alignment arguments prefer
one channel over the other.  We also discuss the possibility that the strongly
coupled gauge interaction can produce multifermion condensates that preserve
the gauge symmetry.  Finally, we show that there are no asymptotically free,
anomaly-free SU($N$) $S_k \, \bar F$ chiral gauge theories with $k \ge 3$,
where $S_k$ denotes the rank-$k$ symmetric representation.  We restrict our
consideration here to chiral gauge theories with only gauge and fermion fields,
but without any scalar fields; the nonperturbative behavior of systems with
interacting gauge, fermion, and scalar fields has been studied, e.g., in
\cite{ghf}, and some recent work on RG flows in chiral theories with scalar
fields includes \cite{srgt}.

This paper is organized as follows.  In Sect. \ref{methods} we briefly review
the theoretical methods that we use for our work. In
Sect. \ref{akfbar_theories} we discuss the construction of the $A_k \, \bar F$
chiral gauge theories and determine the constraints from anomaly cancellation
and asymptotic freedom.  In Sect.  \ref{akfbar_beta_2loop} we ascertain whether
or not the maximal scheme-independent information from the beta function
indicates that the theory has an infrared zero, and, if so, we calculate the
value of $\alpha$ at this zero in the beta function. Sections \ref{gfl_section}
and \ref{mac_section} contain general discussions of the global flavor symmetry
group and the most attractive channel for bilinear fermion condensation
formation in an $A_k \, \bar F$ theory.  In Sects. \ref{a2fbar_theories},
\ref{a3fbar_theories}, and \ref{a4fbar_theories} we present our results on the
specific $A_2 \, \bar F$, $A_3 \, \bar F$, and $A_4 \, \bar F$ classes of 
theories, respectively.  
In Sect. \ref{multifermion_condensates} we discuss multifermion condensates
that can preserve chiral gauge symmetry.  In Sect. \ref{no_skfbar_kge3} we
prove that there are no asymptotically free $S_k \, \bar F$ chiral gauge
theories with $k \ge 3$, where $S_k$ denotes the symmetric rank-$k$ tensor
representation of SU($N$).  Our conclusions are given in
Sect. \ref{conclusions} and some auxiliary formulas are included in two
appendices.


\section{Methods of Analysis}
\label{methods}

In this section we briefly discuss the methods of analysis that we use for our
work. We refer the reader to \cite{cgt,cgt2,cgtsab} for more detailed
discussions of these methods. To determine the constraints due to the
requirement of asymptotic freedom and, for asymptotically free theories, to
study the UV to IR evolution, we calculate the beta function to its maximal
order, namely two-loops. We denote $\alpha(\mu) = g(\mu)^2/(4\pi)$ and $a(\mu)
\equiv g(\mu)^2/16\pi^2$. The beta function is $\beta_\alpha = d\alpha/dt$,
where $dt = d\ln \mu$, with the series expansion
\beq
\beta_\alpha = -2\alpha \sum_{\ell=1}^\infty b_\ell \, a^\ell =
-2\alpha \sum_{\ell=1}^\infty \bar b_\ell \, \alpha^\ell \ . 
\label{beta}
\eeq
In Eq. (\ref{beta}) we have extracted an overall minus sign, $b_\ell$ is the
$\ell$-loop coefficient, and $\bar b_\ell = b_\ell/(4\pi)^\ell$ is the reduced
$\ell$-loop coefficient.  The $n$-loop beta
function, denoted $\beta_{\alpha,n\ell}$, is given by Eq. (\ref{beta}) with the
upper limit on the $\ell$-loop summation equal to $n$ instead of $\infty$. The
requirement of asymptotic freedom means that $\beta_\alpha < 0$ for small
$\alpha$, which holds if $b_1 > 0$.  General expressions for 
$b_1$ \cite{b1} and $b_2$ \cite{b2} are given in Appendix \ref{beta_function}.

Given that $b_1 > 0$, it follows that if $b_2 < 0$, then the two-loop beta
function, $\beta_{\alpha,2\ell}$, has an IR zero at $a_{_{IR,2\ell}}=-b_1/b_2$,
or equivalently,
\beq
\alpha_{IR,2\ell} = -\frac{\bar b_1}{\bar b_2} = -\frac{4\pi b_1}{b_2} \ .
\label{alfir_2loop}
\eeq
For sufficiently large fermion content in an asymptotically free theory, $b_2$
may be negative, so that the beta function exhibits such an infrared zero.
This was discussed for vectorial gauge theories in \cite{b2,bz} and is also
important for the chiral gauge theories under consideration here.  As the
fermion content is reduced, $\alpha_{IR,2\ell}$ increases, and for sufficiently
small fermion content, the beta function may not exhibit any IR zero.  In both
the case where the IR zero occurs at a substantial value of $\alpha_{IR,2\ell}$
and the case where the beta function has no IR zero, the theory becomes
strongly coupled in the infrared.  Higher-loop calculations of the IR zero 
for various fermion representations, including rank-2 tensor representations,
were presented in \cite{bvh,bc}. Since these higher-loop calculations are
scheme-dependent, it was necessary to assess the sensitivity of the IR zero to
scheme transformations. This was done in \cite{sch}-\cite{trschemes}.  For our
purposes here, it will suffice to use the maximal scheme-independent
information available from the beta function, as encoded in the coefficients up
to the two-loop level.

In these situations of strong coupling in the infrared, one can apply various
methods to analyze the resultant nonperturbative behavior of the chiral gauge
theory.  First, one may investigate whether the fermion content of the theory
satisfies the 't Hooft global anomaly matching conditions
\cite{thooft79}. These are necessary but not sufficient conditions for the
gauge interaction to confine and produce massless gauge-singlet composite
spin-1/2 fermions.  If such massless spin-1/2 fermions are actually 
produced, they would saturate the massless-fermion sector of the theory, 
since any composite fermions with spins $J \ge 3/2$ would be massive 
\cite{ww_spinhalf}. 

An alternative possibility is that the gauge interaction can produce bilinear
fermion condensates.  In an irreducibly chiral theory (without a vectorlike
subsector), these condensates break the gauge symmetry, as well as
global flavor symmetries. A common method to identify the most likely channel
in which this condensation occurs is the most-attractive-channel (MAC)
criterion \cite{drs}. Let us consider a fermion condensation channel in which
fermions in the representations $R_1$ and $R_2$ of the gauge group $G$ form a
(Lorentz-invariant) bilinear fermion condensate that transforms according
to the representation $R_c$ of $G$, denoted as
\beq
R_1 \times R_2 \to R_c \ ,
\label{channel}
\eeq
where the subscript $c$ stands for ``condensate channel''.  An
approximate measure of the attractiveness of this condensation channel, is
\beq
\Delta C_2 \equiv C_2(R_1) + C_2(R_2) -C_2(R_c) \ ,
\label{deltac2}
\eeq
where $C_2(R)$ is the quadratic Casimir invariant for the representation $R$
(see Appendix \ref{group_invariants}). In this approach, the most attractive
channel for bilinear fermion condensation is the one with the largest
(positive) value of $\Delta C_2$, and this is thus the most likely to occur. If
two or more such channels have the same value of $\Delta C_2$, then we make use
of a vacuum alignment argument \cite{weinberg76,vacalign}, as follows.  Let us
consider the case where two channels have the same $\Delta C_2$ and produce
condensates in the representations $R_{c_1}$ and $R_{c_2}$.  Assume that these
condensates break the initial gauge group $G$ to the respective subgroups
$H_{c_1} \subset G$ and $H_{c_2} \subset G$. The vacuum alignment argument
favors the condensation channel that yields the larger residual subgroup,
$H_{c_1}$ or $H_{c_2}$, namely the one with the larger order, $o(H_{c_1})$ or
$o(H_{c_2})$ (where the order, $o(H)$, of a Lie group $H$ is the number of
generators of the associated Lie algebra).  This is based on an energy
minimization argument, since the channel that respects the largest residual
gauge symmetry will minimize the number of gauge bosons that pick up masses.

  A rough estimate of the minimal critical strength of the
coupling for fermion condensation in a given channel has been obtained from an
analysis of the Schwinger-Dyson equation for the fermion propagator and is
 \cite{chipt}
\beq
\alpha_{cr,c} \sim \frac{2\pi}{3 \Delta C_2(R_c)} \ .
\label{alfcrit}
\eeq
Owing to the uncertainties inherent in the strong-coupling physics describing
this condensation phenomenon, Eq. (\ref{alfcrit}) is only a rough estimate.
For our purposes, it will be convenient to define the ratio for the channel
(\ref{channel}): 
\beq
\rho_c \equiv \frac{\alpha_{IR,2\ell}}{\alpha_{cr,c}} \ . 
\label{rho}
\eeq
If $\rho_c$ is considerably larger (smaller) than unity, then condensation in
the given channel (\ref{channel}) is likely (unlikely).  We note that fermion
condensation in a strongly coupled gauge theory may, in principle, involve a
product of an even number of fermion operators larger than just two
\cite{lacaze,su2an}.  A conjecture for a thermally motivated inequality
concerning a measure of field degrees of freedom, as evaluated in the UV and in
the IR, was proposed and studied for vectorial and chiral gauge theories in
\cite{dfvgt} and \cite{dfcgt} and investigated further in several works,
including \cite{ads,cgt,cgt2,qed3}.  Here our main methods will consist of
analyses of beta functions, various channels for fermion condensation, and
construction of low-energy effective field theories resulting from
self-breaking of chiral gauge theories.


\section{$A_k \, \bar F$ Theories and Constraints from Anomaly Cancellation and
Asymptotic Freedom}
\label{akfbar_theories}

The chiral gauge theories that we study here have an SU($N$) gauge group and
chiral fermions transforming according to a rank-$k$ antisymmetric tensor
representation $A_k \equiv [k]_N$ of this group, and
the requisite number of chiral fermions in the conjugate fundamental
representation, $\bar F \equiv \overline{[1]}_N$, to render the theories free
of any anomaly in gauged currents \cite{conjugates}.  Here we determine the
constraints on these theories from anomaly cancellation and asymptotic
freedom. These theories are
irreducibly chiral, i.e., they do not contain any vectorlike subsector.
Consequently, the chiral gauge symmetry forbids any fermion mass terms in the
underlying lagrangian.  We denote the number of copies (flavors) of $\bar F$
fermions as $n_{\bar F}$.  The contribution to the triangle anomaly in gauged
currents of a chiral fermion in the $A_k$ representation 
is \cite{anomtk} (see Appendix \ref{group_invariants} )
\beq
{\cal A}([k]_N) = \frac{(N-3)! \, (N-2k)}{(N-k-1)! \, (k-1)!} \ . 
\label{anomak}
\eeq
The total anomaly in the theory is
\beqs
{\cal A} & = & {\cal A}([k]_N)+n_{\bar F} {\cal A}(\overline{[1]}_N) \cr\cr
         & = & {\cal A}([k]_N)-n_{\bar F} {\cal A}([1]_N) \ , 
\label{anomeq}
\eeqs
so ${\cal A}=0$, i.e., the theory is free of anomalies in gauged currents, 
if and only if 
\beq
n_{\bar F} = {\cal A}([k]_N) \ . 
\label{nfbar}
\eeq
If $N$ is even and $k=N/2$, the $[k]_N=[k]_{2k}$ representation is
self-conjugate, with zero anomaly, so Eq. (\ref{nfbar}) yields
$n_{\bar F}=0$ and a nonchiral theory. In order to get a chiral theory, with
positive $n_{\bar F}$, it is necessary and sufficient that
\beq
N \ge N_{min} = 2k+1 \ ,
\label{nmin}
\eeq
so, for a given $k$, we will restrict $N$ to this range.  For $N$ in this
range, the anomaly ${\cal A}([k]_N)$ is an integer greater than unity.  A
member of this set of chiral gauge theories is thus determined by its values of
$k$ and $N$ and has the form
\beq
G={\rm SU}(N), \ {\rm fermions}: \ A_k + n_{\bar F} \, \bar F \ , 
\label{akfbar_fermions}
\eeq
i.e., $[k]_N+n_{\bar F}\overline{[1]}_N$, 
where $N$ is bounded below by (\ref{nmin}). The $A_k \, \bar F$ theories
with $k=3$ and $k=4$ have respective upper bounds on $N$ imposed by the
requirement of asymptotic freedom.

To determine the upper bounds on $N$ for these values of $k$, we calculate the
one-loop coefficient in the beta function, $b_1$.  To indicate explicitly the
dependence of the $b_\ell$ coefficients with $\ell=1,2$ on $k$, we shall write
them as $b_\ell^{(k)}$.  For a general SU($N$) $A_k \, \bar F$ theory, we have
\beqs
b_1^{(k)} & = &
\frac{1}{3}\bigg [ 11N -2\Big \{ T(A_k) + n_{\bar F} \, T(\bar F) \Big \} 
\bigg ] \cr\cr
& = & \frac{1}{3}\bigg [ 11N - \frac{1}{(k-1)!}\Big \{ 
\Big [\prod_{j=2}^k (N-j) \Big ] \cr\cr
& + & \frac{(N-3)!(N-2k)}{(N-k-1)!} \Big \} \bigg ] \ . 
\label{b1akfbar}
\eeqs
In Eq. (\ref{b1akfbar}), both $T(A_k)$ and $n_{\bar F} = {\cal A}(A_k)$ 
are polynomials of degree ${\rm max}(1,k-1)$ in $N$ and hence, 
$b_1^{(k)}$ is a polynomial of degree 
${\rm max}(1,k-1)$ in $N$. Specifically, we find 
\beq
b_1^{(2)} = 3N+2 \ , 
\label{b1_a2fbar}
\eeq
\beq
b_1^{(3)} = \frac{1}{3}(-N^2+18N-12) \ , 
\label{b1_a3fbar}
\eeq
and
\beq
b_1^{(4)} = \frac{1}{9}(-N^3+12N^2-14N+60) \ . 
\label{b1_a4fbar}
\eeq
The $A_2 \, \bar F$ theories are thus asymptotically free without any upper
bound on $N$. For the $A_3 \, \bar F$ theories, the asymptotic-freedom
requirement that $b_1^{(3)}$ must be positive yields 
the upper bound $N \le 17$.  (If one were to generalize $N$ from
positive integer values to real values, $b_1^{(3)}$ is positive for $N$ in the
range $9-\sqrt{69} < N < 9+\sqrt{69}$, i.e., $0.6934 < N < 17.3066$ to the
indicated floating-point accuracy.) In the $A_4 \, \bar F$ theories, 
$b_1^{(4)}$ is positive only for the integer values $N=9, \ 10, \ 11$. 
(With $N$ generalized to a positive real number, 
$b_1^{(4)} > 0$ for $N < 11.2291$.) Denoting $N_{max}$ as the maximal value of
$N$, for a given $k$, for which an SU($N$) $A_k \bar F$ theory is
asymptotically free, we summarize these results as 
\beq
N_{max} = \cases{ \infty & for $k=2$ \cr
                  17     & for $k=3$ \cr  
                  11     & for $k=4$} \ . 
\label{nmax}
\eeq

Combining these results, we explicitly exhibit the asymptotically free,
anomaly-free chiral gauge theories of this type with 
$2 \le k \le 4$ together with the respective allowed ranges of $N$,
$N_{min} \le N \le N_{max}$:
\beq
k=2 \ \Longrightarrow \ N \ge 5 \ , 
\label{k2_Ninterval}
\eeq
\beq
k=3 \ \Longrightarrow \ 7 \le N \le 17 \ , 
\label{k3_Ninterval}
\eeq
\beq
k=4 \ \Longrightarrow \ 9 \le N \le 11 \ . 
\label{k4_Ninterval}
\eeq

The SU($N$) $A_2 \, \bar F$ theories have been studied in several works
\cite{drs,by,georgi86,dfcgt,ads,cgt}, has fermion content given by
\beq
k=2: \quad {\rm fermions}: A_2 + (N-4) \, \bar F \ ,
\label{a2fbar}
\eeq
The SU($N$) $A_k \, \bar F$ theories with $k=3$ and $k=4$ are, to our 
knowledge, new here.  These have the fermion contents
\beqs
k=3: \quad & & {\rm fermions}: A_3 \ + \ \frac{(N-3)(N-6)}{2} \,
\bar F \cr\cr
& & 
\label{a3fbar}
\eeqs
and
\beqs
k=4: \quad & & {\rm fermions}: 
A_4 \ + \ \frac{(N-3)(N-4)(N-8)}{6} \, \bar F \ . \cr\cr
& & 
\label{a4fbar}
\eeqs

For the $A_k \, \bar F$ theories with $k=2, \ 3, \ 4$, we denote the fermion
field in the $A_k=[k]_N$ representation as $\psi^{ab}_L$, $\psi^{abd}_L$, and
$\psi^{abde}_L$, respectively, where $a,b,d,e$ are SU($N$) gauge indices (the
symbol $c$ is reserved to mean charge conjugation) with $N$ is in the
respective intervals (\ref{a2fbar})-(\ref{a4fbar}), and we denote the $\bar F$
fermions as $\chi_{a,i,L}$, where $i$ is a copy (flavor) index taking values in
the respective ranges $1 \le i \le n_{\bar F} = {\cal A}([k]_N)$.

We next show that there are no asymptotically free $A_k \, \bar F$ theories 
with $k \ge 5$.  Consider first the $k=5$ theory, for which
\beqs
& & b_1^{(5)} = \frac{1}{36}(-N^4+18N^3-119N^2+474N-360) \ . \cr\cr
& & 
\label{b1_ak5fbar}
\eeqs
With $N$ generalized to a real variable, $b_1^{(5)}$ is positive only for
$N$ in the range $0.9585 < N < 10.7379$.  But for an $A_k \, \bar F$ theory, 
$N$ is bounded below by $2k+1$, which has the value 11 here, so for this $k=5$
theory there is no value of $N$ that simultaneously satisfies both the
lower bound (\ref{nmin}) and the requirement of asymptotic freedom. 
We reach the same conclusion in the $k=6$ case, for which 
\beq
b_1^{(6)} = \frac{1}{180}(-N^5+25N^4-245N^3+1175N^2-2094N+2520) \ . 
\label{b1_ak6fbar}
\eeq
With $N$ extended from physical values to real numbers, $b_1^{(6)}$ is positive
if $N < 11.098$, but $N$ is required to satisfy $N \ge 13$, which again means
that for $k=6$ there is no value of $N$ that satisfies the lower bound
(\ref{nmin}) and the requirement of asymptotic freedom.  Similarly, we
find that for the $k=7$ case, $b_1^{(7)}$ is only positive for the range 
$1.094 < N < 11.742$, while $N$ must be in the range $N \ge 15$ by
(\ref{nmin}), and so forth for higher $k$. The underlying reason for the
non-existence of asymptotically free $A_k \, \bar F$ chiral gauge theories with
these higher values of $k$ is that, as noted above, both $T([k]_N)$ and 
${\cal A}([k]_N)$ are polynomials of degree ${\rm max}(1,k-1)$ in $N$, and 
they both contribute negatively to $b_1^{(k)}$ for the relevant range 
$N \ge 2k+1$. Their negative contributions eventually outweigh the positive 
contribution of the $(11/3)N$ term from the gauge fields.

In passing, we remark that there are two possible ways that one could expand
the fermion content of the $A_k \, \bar F$ models considered here for certain
$k$ and $N$ values, as restricted by the constraint of asyamptotic freedom,
namely (i) to have $n_{cp}$ replications of the chiral fermion content and (ii)
to add vectorlike subsectors.  For example, in category (i), the following
$k=3$ theories are asymptotically free: $n_{cp}=2$ and $7 \le N \le 11$;
$n_{cp}=3$ and $7 \le N \le 9$; $n_{cp}=4$ and $N=7, \ 8$; and $n_{cp}=5$ and
$N=7$. We have studied different chiral gauge theories with this sort of
$n_{cp}$ replication of a minimal irreducible chiral fermion content in
\cite{cgtsab}.  We shall not pursue these expansions here but instead focus on
studying the minimal $A_k \, \bar F$ theories.


\section{Beta Function Analysis of $A_k \, \bar F$ Theories}
\label{akfbar_beta_2loop}

In this section we give a general analysis of the beta function applicable to
all of the (anomaly-free) asymptotically free $A_k \, \bar F$ theories, with
$N$ in the respective ranges $N \ge 5$ for $k=2$ and the finite intervals $7
\le N \le 17$ for $k=3$ and $9 \le N \le 11$ for $k=4$ as given in
(\ref{a2fbar})-(\ref{a4fbar}).  In Sect. \ref{akfbar_theories} we gave the
one-loop coefficient for the $A_k \, \bar F$ theories, which we used to
determine the upper bound on $N$ for a given $k$. Here we proceed to give the
two-loop coefficient, $b_2^{(k)}$, and use it to analyze the UV to IR
evolution.  We have (again with $A_k \equiv [k]_N$, and $F \equiv [1]_N$)
\beqs
b_2^{(k)} & = & 
\frac{1}{3}\bigg [ 34N^2-2 \Big \{ \Big (5C_2(G) + 3C_2(A_k)\Big )T(A_k)
\cr\cr
& + &  n_{\bar F} \Big (5C_2(G) + 3C_2(\bar F)\Big )T(\bar F) \Big \} \ 
\bigg ] \ , 
\label{b2_akfbar}
\eeqs
where the various group invariants are listed in appendix
\ref{group_invariants}.  For the three relevant cases, $k=2, \ 3, \ 4$, the
explicit expressions are
\beq
b_2^{(2)} = \frac{13N^3+30N^2+N-12}{2N} \ , 
\label{b2_a2fbar}
\eeq
\beq
b_2^{(3)} = \frac{-16N^4+183N^3-204N^2-27N+108}{6N} \ , 
\label{b2_a3fbar}
\eeq
and
\beqs
b_2^{(4)} & = & (36N)^{-1}\Big [-35N^5+429N^4-1321N^3+2235N^2 \cr\cr
          & + & 588N-1440\Big ] \ . 
\label{b2_a4fbar}
\eeqs

In Table \ref{akfbar_properties} we list values of the reduced coefficients
$\bar b_1$ and $\bar b_2$ for an illustrative set of the $A_2 \, \bar F$
theories and for all of the (asymptotically free) $A_3 \, \bar F$ and 
$A_4 \, \bar F$ theories.  In the cases where $\bar b_2 < 0$ so that the 
two-loop beta function has a physical IR zero, we have also listed the value of
$\alpha_{IR,2\ell}$.  The value of the resultant ratio $\rho_c$ for
condensation in the most attractive channel for bilinear fermion condensation
(discussed further below) gives an estimate of whether the theories are weakly
or strongly coupled in the infrared. This is indicated by the abbreviations WC,
MC, and SC (weak coupling, moderate coupling, and strong coupling) in Table
\ref{akfbar_properties}.


\section{Global Symmetry of $A_k \, \bar F$ Theories}
\label{gfl_section} 

Because the $A_k \, \bar F$ theories are irreducibly chiral, so that the chiral
gauge symmetry requires the fermions to be massless, each such theory has a
classical global flavor symmetry
\beq
G_{fl,cl}^{(k)} = {\rm U}(n_{\bar F})_{\bar F} \otimes {\rm U}(1)_{A_k} \ , 
\label{glcl_akfbar}
\eeq
where $n_{\bar F}={\cal A}([k]_N)$ as given in Eq. (\ref{nfbar}).  
Equivalently, 
\beq
G_{fl,cl}^{(k)} = \cases { {\rm U}(1)_{\bar F} \otimes {\rm U}(1)_{A_k} & if 
$n_{\bar F}=1$ \cr
  {\rm SU}(n_{\bar F})_{\bar F} \otimes {\rm U}(1)_{\bar F} \otimes 
{\rm U}(1)_{A_k} & if $n_{\bar F} \ge 2$}
\label{glcl_akfbar2}
\eeq
For $n_{\bar F} \ge 2$, the multiplet
$(\chi_{a,1,L},...,\chi_{a,n_{\bar F},L})$ may be taken to transform as the 
conjugate fundamental, $\overline{\fund}$, representation of the global 
flavor group, SU($n_{\bar F}$). 
The U(1)$_{\bar F}$ and U(1)$_{A_k}$ symmetries in (\ref{glcl_akfbar2}) are
both broken by SU($N$) instantons \cite{instantons}. As in
\cite{cgt}, we define a vector whose components are comprised of
the instanton-generated contributions to the breaking of these symmetries.
In the basis $(A_k,\bar F)$, this vector is
\beqs
{\vec v}^{(k)} & = & \Big ( T([k]_N), \ n_{\bar F} T(\bar F) \Big ) \cr\cr
& = & \lambda_{N,k} (N-2,N-2k) \ , 
\label{anomvec}
\eeqs
where
\beq
\lambda_{N,k}= \frac{(N-3)!}{2(k-1)!(N-k-1)!} \ . 
\label{lambda}
\eeq
We can construct one linear combination of the two original currents that is
conserved in the presence of SU($N$) instantons. We denote the corresponding
global U(1) flavor symmetry as U(1)$^\prime$ and the fermion charges under this
U(1)$^\prime$ as
\beq
{\vec Q}^{(k) \prime} = \Big ( Q'_{A_k}, \ Q'_{\bar F} \Big ) \ .
\label{qvec}
\eeq
The U(1)$^\prime$ current is conserved if and only if
\beq
\sum_f n_f T(R_f) \, Q_f^{(k) \prime} = 
{\vec v} \cdot {\vec Q}^{(k) \prime} = 0 \ .
\label{u1inv}
\eeq
This condition only determines the vector ${\vec Q}^{(k) \prime}$ up
to an overall multiplicative constant.  A solution is
\beq
{\vec Q}^{(k) \prime} = \Big ( N-2k, \ -(N-2) \Big ) \ .
\label{qvecsol}
\eeq
The actual global chiral flavor symmetry group (preserved in the presence of
instantons) is then
\beq
G_{fl}^{(k)} = \cases{ {\rm U}(1)'  & if $n_{\bar F}=1$ \cr
   {\rm SU}(n_{\bar F}) \otimes  {\rm U}(1)' & if $n_{\bar F} \ge 2$} \ . 
\label{gfl}
\eeq
For the three $k$ values relevant here, this is 
\beq
G_{fl}^{(2)} = \cases { {\rm U}(1)' & if $N=5$ \cr
   {\rm SU}(N-4)_{\bar F} \otimes {\rm U}(1)' & if $N \ge 6$ } 
\label{gflk2}
\eeq
with U(1)$^\prime$ charges 
\beq
{\vec Q}^{(2) \prime} = \Big ( N-4, \ -(N-2) \Big ) \ ,
\label{qveck2}
\eeq
\beq
G_{fl}^{(3)} = {\rm SU}\Big ( \frac{(N-3)(N-6)}{2} \Big )_{\bar F} 
\otimes {\rm U}(1)'
\label{gflk3}
\eeq
with U(1)$^\prime$ charges 
\beq
{\vec Q}^{(3) \prime} = \Big ( N-6, \ -(N-2) \Big ) \ ,
\label{qveck3}
\eeq
and
\beq
G_{fl}^{(4)} = {\rm SU}\Big ( \frac{(N-3)(N-4)(N-8)}{6} \Big )_{\bar F}
\otimes {\rm U}(1)'
\label{gflk4}
\eeq
with U(1)$^\prime$ charges 
\beq
{\vec Q}^{(4) \prime} = \Big ( N-8, \ -(N-2) \Big ) \ .
\label{qveck4}
\eeq
%


\section{Most Attractive Channel for Bilinear Fermion Condensation in 
$A_k \, \bar F$ Theories}
\label{mac_section}

\subsection{General Analysis} 

The ultraviolet to infrared evolution of a particular SU($N$) $A_k \, \bar F$
theory is determined by the values of $N$ and $k$.  In the cases where it can
lead to the formation of a bilinear fermion condensate, one should then
determine the most attractive channel in which this condensate can form.  We
present this analysis here.  Since the $A_k \, \bar F$ theories that we
consider here are irreducibly chiral, a bilinear condensate breaks the gauge
symmetry.  In Sect. \ref{multifermion_condensates} below, we will discuss the
possible formation of multifermion condensates involving more than just two
fermions, which can preserve the chiral gauge symmetry.

For the theories that we are discussing here, there are two relevant bilinear
fermion condensation channels. First, there is a channel with a condensate that
involves the contraction of $2k$ gauge indices of the antisymmetric tensor 
density $\epsilon_{a_1,...,a_N}$ with the bilinear fermion product 
$A_k \times A_k$, which transforms like $\bar A_{N-2k}$. This channel can 
thus be written as 
\beq
A_k \times A_k \to \bar A_{N-2k} \ . 
\label{akak_to_anminus2k_channel}
\eeq
This channel has attractiveness measure 
\beq
\Delta C_2 = \frac{k^2(N+1)}{N} \quad {\rm for} \ 
A_k \times A_k \to \bar A_{N-2k} \ . 
\label{deltac2_akak_to_anminus2k_channel}
\eeq
For a given $k$, this $\Delta C_2$
is a monotonically decreasing function of $N$, decreasing gradually from its
value at $N=2k+1$,
\beqs
\Delta C_2 & = & \frac{2k^2(k+1)}{2k+1} = \frac{(N-1)^2(N+1)}{4N} \quad 
{\rm at} \ N=2k+1 \cr\cr
& & {\rm for} \ A_k \times A_k \to \bar A_{N-2k} = \bar A_1 = \bar F 
\label{deltac2_akak_Nmin}
\eeqs
and approaching the limit $k^2$ for $N \gg k$.

Second, there is the channel 
\beq
A_k \times \bar F \to A_{k-1} \ , 
\label{akfbar_to_akminus1_channel}
\eeq
with
\beq
\Delta C_2 = \frac{(N+1)(N-k)}{N} \quad {\rm for} \ 
A_k \times \bar F \to A_{k-1} \ .
\label{deltac2_akfbar_to_akminus1_channel}
\eeq
For a given $k$, this $\Delta C_2$ is a monotonically increasing function of
$N$, increasing from the value 
\beqs
\Delta C_2 & = & \frac{2(k+1)^2}{2k+1} = \frac{(N+1)^2}{2N} \quad 
{\rm at} \ N=2k+1 \cr\cr
& & {\rm for} \ A_k \times \bar F \to A_{k-1} \ ,
\label{deltac2_akfbar_Nmin}
\eeqs
and approaching a linear growth with $N$ for $N \gg k$.  In Table
\ref{kN_DeltaC2_values} we list the value of $\Delta C_2$ in
Eq. (\ref{deltac2_akak_to_anminus2k_channel}) for the 
$A_k \times A_k \to \bar A_{N-2k}$ channel and the value of $\Delta C_2$ in
Eq. (\ref{deltac2_akfbar_to_akminus1_channel}) for the 
$A_k \times \bar F \to A_{k-1}$ channel for an illustrative set of 
$A_2 \, \bar F$ theories and for
the full set of (asymptotically free) $A_3 \, \bar F$ and $A_4 \, \bar F$
theories.

The most attractive channel for bilinear fermion condensation is the one among
these two channels with the larger value of $\Delta C_2$ (assuming that these
two values are unequal; we discuss the cases where they are equal below).  For
a given value of $k$, we thus determine the MAC as a function of $N$ in its
allowed range $N_{min} \le N \le N_{max}$ by examining the difference,
\beqs
& & \Delta C_2(A_k \times A_k \to \bar A_{N-2k}) - 
\Delta C_2(A_k \times \bar F \to A_{k-1}) \cr\cr
& = & \Big ( \frac{N+1}{N}\Big ) \Big [ k(k+1)-N \Big ] \ . 
\label{deltac2_diff}
\eeqs
For $N=N_{min}=2k+1$, $\Delta C_2$ is larger for the first channel, $A_k \times
A_k \to \bar A_{N-2k}= \bar A_1 = \bar F$, than for the second channel, 
$A_k \times \bar F \to A_{k-1}$.  This is evident analytically from the fact 
that with $N=2k+1$, the difference (\ref{deltac2_diff}) is
\beq
\frac{2(k+1)(k^2-k-1)}{2k+1} = \frac{(N+1)(N^2-4N-1)}{4N} \ , 
\label{deltac2_diff_Nmin}
\eeq
which is positive for the relevant range $k \ge 2$ considered here. 
Since $\Delta C_2$ for the first channel
decreases monotonically as a function of $N$, while the $\Delta C_2$ for the
second channel increases monotonically as a function of $N$, it follows that at
some value of $N$, which we denote $N_e$ (where $e$ stands for ``equal''),
these values are equal, and for $N > N_e$, the $\Delta C_2$ for the second
channel is larger than that for the first channel.  Setting the two $\Delta
C_2$ values equal and solving for $N=N_e$, we find
\beq
N_e = k(k+1) \ .
\label{ne}
\eeq
Evaluating Eq. (\ref{ne}) for the three relevant values of $k$, we have
\beq
N_e = \cases{ 6   & for $k=2$ \cr
              12  & for $k=3$ \cr
              20  & for $k=4$} \ . 
\label{nevalues}
\eeq
The first two of these values are within the respective allowed ranges for $N$,
while the value for $k=4$ is larger than the upper bound $N_{max}=11$ for
$k=4$. 

Consequently, with $N_{min}=2k+1$ and $N_{max}$ as given in Eqs. 
(\ref{nmin}) and (\ref{nmax}), we find that, for a given $k$, 
\beqs
& & {\rm If} \ 2k+1 \le N < k(k+1) \ {\rm then} \cr\cr
& & {\rm MAC} \ = \ A_k \times A_k \to \bar A_{N-2k} \cr\cr
& & {\rm If} \ k(k+1) < N \le N_{max} \ {\rm then} \cr\cr
& & {\rm MAC} \ = \ A_k \times \bar F \to A_{k-1} \ , 
\label{akfbarmac}
\eeqs
with the proviso that the second possibility only applies if 
$k(k+1) < N_{max}$, and hence only for $k=2$ and $k=3$.  
Thus, in particular, if $N=N_{min}=2k+1$, then the MAC is the special case of
(\ref{akak_to_anminus2k_channel}):
\beq
{\rm If} \ N=2k+1 \ {\rm then \ MAC} = A_k \times A_k \to \bar F \ . 
\label{akak_to_fbar_channel}
\eeq
In addition to breaking the original SU($N$) gauge symmetry, these condensates
also break both the non-Abelian factor group SU($n_{\bar F}$) (which is present
if $n_{\bar F} \ge 2$) and the U(1)$'$ factor group in the global flavor
symmetry (\ref{gfl}).  In particular, the breaking of the U(1)$'$ symmetry is
evident from the fact that the respective condensates in these channels have 
the nonzero U(1)$'$ charges
\beq
Q'^{(k)} = 2Q'_{A_k}=2(N-2k) \quad {\rm for} \ A_k \times A_k \to \bar 
A_{N-2k}
\label{q_akak_channel}
\eeq
and
\beq
Q'^{(k)} = Q'_{A_k} + Q'_{\bar F} = 2(1-k) \quad {\rm for} \ 
A_k \times \bar F \to A_{k-1} \ . 
\label{q_akfbar_channel}
\eeq

The marginal case $N=N_e=k(k+1)$ requires further analysis, since the $\Delta
C_2$ values for the $A_k \times A_k \to \bar A_{N-2k}$ and 
$A_k \times \bar F \to A_{k-1}$ channels are equal, so the procedure of 
picking the channel
with the largest $\Delta C_2$ cannot determine which is more likely to occur.
To deal with this marginal case, we use a vacuum alignment argument, which, as
applied to possible bilinear fermion condensation channels, favors the one
whose condensate respects the larger residual gauge symmetry. To apply the
vacuum alignment argument, we must thus determine the residual gauge symmetry
group respected by the condensates that occur in these two channels. The
resultant bilinear fermion condensate transforms like an $n$-fold antisymmetric
tensor representation of SU($N$), where $n=N-2k$ for the 
$A_k \times A_k \to \bar A_{N-2k}$ channel and $n=k-1$ for the 
$A_k \times \bar F \to A_{k-1}$ channel.  (The fact that in the first case the
condensate transforms like $\bar A_{N-2k}$ rather than $A_{N-2k}$ does not
affect how this breaks SU($N$).) 
From the point of view of the group theory, the problem of
determining the residual gauge symmetry is effectively the same as the problem
of determining the residual gauge symmetry that results when one has a Higgs
field transforming according to the antisymmetric rank-$n$ representation of
SU($N$).  An analysis of this, within the context of Higgs-induced symmetry
breaking, was given in \cite{kim82}, and the results depend, in that context,
on the parameters in the Higgs potential, which one has the freedom to choose,
subject to the overall constraint that the energy must be bounded below.  As
emphasized in Ref. \cite{isb}, the situation is different in dynamical gauge
symmetry breaking; in principle, given an initial gauge group and set of
fermions, there is a unique answer for how the symmetry breaks; this breaking
does not depend on any parameters in a Higgs potential.  Despite this basic
difference between dynamical and Higgs-induced gauge symmetry breaking, we can
make use of the general group-theoretic analysis performed for the Higgs
case. The result is that there are, {\it a priori}, three possibilities for the
gauge symmetries respected by a condensate or Higgs vacuum expectation value
transforming as the rank-$n$ antisymmetric tensor representation of SU($N$),
$[n]_N$. Denoting the integral part of a real number $r$ as $[r]$ and setting
\beq
\kappa \equiv [N/n] \ , 
\label{kappa}
\eeq
these are \cite{kim82} 
\beq
{\rm SU}(N-n) \otimes {\rm SU}(n) \quad {\rm with} \ 2 \le n < [N/2] \ , 
\label{h_susu}
\eeq
\beqs
& & [{\rm SU}(n)]^\kappa \quad {\rm with} \ 3 \le n \le [N/2] \ {\rm if} \ 
N-[N/n]n=0 \ {\rm or} \ 1 \ , \cr\cr
& & 
\label{h_sukappa}
\eeqs
and the symplectic group
\beq
{\rm Sp}(2\kappa) \quad {\rm if} \ n=2 \ .  
\label{h_sp}
\eeq
We analyze the respective cases $k=2, \ 3, \  4$ next. 


\subsection{Case $k=2$ }

From the special case for $k=2$ of our general result (\ref{akfbarmac}) above,
we infer that the $A_2 \times A_2 \to \bar A_1 = \bar F$ channel is the most
attractive channel for bilinear fermion condensation in the $A_2 \, \bar F$
theories for the lowest value of $N$, namely $N=5$, while the $A_2 \times \bar
F \to F$ channel is the MAC for the infinite interval $N \ge 7$.  For the
marginal case $k=2$, $N=6$, the $A_2 \times A_2 \to \bar A_{N-2k}=\bar A_2$ and
$A_2 \times \bar F \to F$ channels have the same value of $\Delta C_2$, namely
$\Delta C_2=14/3=4.667$ (see Table \ref{kN_DeltaC2_values}), so the 
$\Delta C_2$ attractiveness criterion cannot be used to decide which is more 
likely to occur.  Now the
condensate in the $A_2 \times \bar F \to F$ channel leaves invariant an SU(5)
subgroup of SU(6), with order 24.  To analyze the possible invariance groups of
a condensate in the $A_2 \times A_2 \to \bar A_2$ channel, we apply our
discussion above with $N=6$, $n=2$, and hence $\kappa=[6/2]=3$, so the 
{\it a  priori} possible invariance groups of the condensate are 
${\rm SU}(4) \otimes {\rm SU}(2)$ with order 18 and ${\rm Sp}(6)$ with order 
21.  Neither of these
groups has an order as large as that of SU(5), so the vacuum alignment argument
predicts that, if a bilinear fermion condensate forms, then this condensate 
will form in the $A_2 \times \bar F \to F$ channel.  Summarizing our results 
for $k=2$ and all $N$, we thus find that if bilinear fermion condensation 
occurs, then
\beq
k=2 \ \Longrightarrow {\rm MAC} = \cases{ A_2 \times A_2 \to \bar A_1 
& for $N=5$ \cr
  A_2 \times \bar F \to F & for $N \ge 6$ } \ . 
\label{k2mac}
\eeq

As noted above, since this class of (asymptotically free) $A_2 \, \bar F$
chiral gauge theories satisfies the 't Hooft global anomaly matching
conditions, there is also the possibility of confinement, yielding massless
composite fermions.  There is also the possibility of multifermion condensate
formation, which we will discuss below. Since the
early works such as \cite{thooft79,drs,by}, for a class of asymptotically free
chiral gauge theories such as the $A_2 \, \bar F$ class discussed here, for
which the UV to IR evolution leads to strong coupling and hence 
could lead to confinement with massless composite
fermions or to fermion condensation, there has not, to our knowledge, been a
rigorous argument presented that actually determines the type of UV to IR 
evolution in an asymptotically free chiral gauge theory. 


\subsection{Case $k=3$}

From the special case for $k=3$ of our general result (\ref{akfbarmac}) above,
we infer that the $A_3 \times A_3 \to \bar A_{N-2k}= \bar A_{N-6}$ channel is
the most attractive channel for bilinear fermion condensation not only for the
minimal value of $N$, namely $N=7$, but also for the interval of $N$ values up
to $N=11$.  We discuss the marginal case of $N=12$ last.  Again substituting
$k=3$ into (\ref{akfbarmac}), it follows formally that the MAC for 
$12 \le N \le 17$ is the $A_3 \times \bar F \to A_2$ channel.  However, for 
$N=13, \ 14$,
the respective values of the IR zero in the beta function are sufficiently
close to the rough estimate of the minimal critical value of $\alpha$ for
condensate formation in the $A_3 \times \bar F \to A_2$ channel (see Tables
\ref{akfbar_properties} and \ref{kN_DeltaC2_values}) that it is possible that
the system could evolve from the UV to a deconfined, non-Abelian Coulomb phase
in the IR with no fermion condensate formation or associated spontaneous chiral
symmetry breaking.

The $k=3$, $N=12$ case is again marginal; the $A_3 \times A_3 \to \bar A_6$ and
$A_3 \times \bar F \to A_2$ channels have the same value of $\Delta C_2$,
namely $\Delta C_2=39/4=9.750$.  Hence, we use a vacuum alignment argument to
decide on which of these channels is more likely to occur.  For the $A_3 \times
A_3 \to \bar A_{N-6}=\bar A_6$ channel, we apply our discussion above with
$N=12$, $n=6$, and hence $\kappa=[12/6]=2$, so the invariance group of the
$\bar A_2$ condensate is $[{\rm SU}(6)]^2$, with order 70.  For the 
$A_3 \times \bar F \to A_2$ channel, we have $N=12$, $n=2$ and hence 
$\kappa=[12/2]=6$, so
the {\it a priori} possible invariance groups of the $A_2$ condensate are 
${\rm SU}(10) \otimes {\rm SU}(2)$ with order 102 and ${\rm Sp}(12)$ with 
order 78. The vacuum alignment argument thus favors condensation in the 
$A_3 \times \bar F \to \bar A_2$ channel for this $N=12$ case. 
Summarizing these results, we have
\beqs
& & k=3 \ \Longrightarrow {\rm MAC}= \cases{ A_3 \times A_3 \to \bar A_{N-6} & 
for $7 \le N \le 11$ \cr
A_3 \times \bar F \to A_2 & for $12 \le N \le 17$ } \cr\cr
& & 
\label{k3mac}
\eeqs
However, as mentioned above, for $N=13, \ 14$ (and also for $N=12$), the
respective values of $\rho_c$ are sufficiently close to unity that, in view of
the intrinsic theoretical uncertainties in the analysis of the strong-coupling
physics, it is possible that the UV to IR evolution could lead either to the
formation of a fermion condensate or to a non-Abelian Coulomb phase without
spontaneous chiral symmetry breaking.

If $N$ is in the higher interval $15 \le N \le 17$, then
$\rho_c$ is sufficiently small that we definitely expect the evolution to lead
to a chirally symmetric non-Abelian Coulomb phase in the IR.  Hence, in these
cases, the MAC is not directly relevant to the dynamics of the theory.


\subsection{Case $k=4$} 

Finally, we discuss the theories with $k=4$, for which the interval of values
of $N$ is $9 \le N \le 11$.  Since the value of $N_e$, namely $N_e=20$, is
larger than $N_{max}$, the most attractive channel for bilinear fermion
condensation in all of these theories is $A_4 \times A_4 \to \bar A_{N-8}$,
i.e., $A_4 \times A_4 \to \bar F$ for $N=9$, $A_4 \times A_4 \to \bar A_2$ for
$N=10$, and $A_4 \times A_4 \to \bar A_3$ for $N=11$. In the SU(9) $A_4 \, \bar
F$ theory, the IR zero in the two-loop beta function is much larger than
$\alpha_{cr}$ for this channel, so it is likely that the SU(9) gauge
interaction would produce a condensate in this channel, thereby breaking SU(9)
to SU(8).  For $N=10$, $\alpha_{IR,2\ell}/\alpha_{cr} = 1.7$, which is
sufficiently close to unity that, taking account of the uncertainties in the
strong-coupling estimates, the UV to IR evolution might produce a condensate in
the respective most attractive bilinear fermion channel or might lead to a
non-Abelian Coulomb phase. For $N=11$, the IR zero in the two-loop beta
function is small compared with the estimated $\alpha_{cr}$ for the $A_4 \times
A_4 \to \bar A_3$ condensation channel, so we definitely expect the system to
evolve from the UV to a non-Abelian Coulomb phase in the IR.


\section{$A_2 \, \bar F$ Theories} 
\label{a2fbar_theories}

\subsection{General}

In this section we analyze the UV to IR evolution of some $A_2 \, \bar F$
theories in detail.  Recall that the explicit fermion fields are 
$A_2: \ \psi^{ab}_L$ and $\bar F: \ \chi_{a,i,L}$, where $a,\ b$ are the
SU($N$) gauge indices and $i=1,..,N-4$ is a copy (flavor) index. 
The one-loop and two-loop coefficients were given in
Eqs. (\ref{b1_a2fbar}) and (\ref{b2_a2fbar}).  We find that for all 
$N \ge N_{min}=5$, the coefficient $b_2^{(2)}$ is positive, so the two-loop 
beta function of the $A_2 \, \bar F$ theory has no IR zero. Hence, 
as the Euclidean reference scale $\mu$ decreases from the UV to the IR, the
gauge coupling increases until it eventually exceeds the region where it is
perturbatively calculable. This IR behavior is thus marked as SC, for strong
coupling, in Table \ref{akfbar_properties}. 

The global flavor symmetry group for this theory is given in Eq. (\ref{gflk2})
with the U(1)$^\prime$ charge assignments in (\ref{qveck2}).  This theory
satisfies the 't Hooft global anomaly matching conditions
\cite{thooft79,dfcgt}, so, as it becomes strongly coupled in the infrared, it
could confine and produce massless gauge-singlet composite spin-1/2 fermions as
well as massive gauge-singlet mesons and also primarily gluonic states.
If this happens, then it is a complete
description of the UV to IR evolution. The three-fermion operator for the
composite gauge-singlet fermion can be written as 
\beq
f_{ij} \propto [\chi_{a,i,L}^T C \psi^{ab}_L] \chi_{b,j,L} + (i \leftrightarrow
j) \ . 
\label{fij}
\eeq
From Eq. (\ref{qveck2}), the U(1)$'$ charge of this composite fermion is
\beq
Q_{f_{ij}} = Q_{\bar A} + 2Q_{\bar F} = -N \ . 
\label{qfij}
\eeq
If $N \ge 6$, $f_{ij}$ transforms as the conjugate symmetric rank-2 tensor
representation, $\overline{\sym}$, of the ${\rm SU}(N-4)_{\bar F}$ factor group
in the global flavor symmetry group 
$G_{fl}^{(2)}={\rm SU}(N-4)_{\bar F} \otimes {\rm U}(1)'$ of the theory.

Another possibility is that the SU($N$) gauge interaction could produce
bilinear fermion condensates, thereby breaking both gauge and global
symmetries.  The most attractive channel for this fermion condensation was
determined, as a function of $N$, in Eq. (\ref{k2mac}).  It can also be
possible to form multifermion condensates involving more than two fermion
fields, which preserve the chiral gauge symmetry. 
We will discuss this latter possibility in
Sect. \ref{multifermion_condensates}.  Here we proceed to analyze bilinear
fermion condensate formation for various specific theories.  


\subsection{SU(5) $A_2 \, \bar F$ Theory}

The simplest chiral gauge theory in the $A_2 \, \bar F$ family of theories has
the gauge group SU(5), with fermion content given by the $N=5$ special case of 
Eq. (\ref{a2fbar}), namely $A_2+\bar F = [2]_5 + \overline{[1]}_5$. 
Like the other $A_2 \, \bar F$ theories considered here that become strongly
coupled in the infrared, this one could confine and produce a massless
composite fermion.  Alternatively, it could produce fermion condensates. 
The most attractive channel for bilinear fermion condensation in 
this theory is $A_2 \times A_2 \to \bar A_1$.  If the dynamics is such that
this condensate does, indeed, form, then we denote the mass scale at which it 
is produced as $\Lambda_5$.  This condensate breaks the
SU(5) gauge symmetry to SU(4). Without loss of generality, we take the gauge
index corresponding to the breaking direction to be $a=5$.  The condensate then
has the form
\beqs
& & \langle \epsilon_{abde5} \psi^{ab \ T}_L C \psi^{de}_L\rangle \propto
\Big [ \langle \psi^{12 \ T}_L C \psi^{34}_L\rangle
       -\langle \psi^{13 \ T}_L C \psi^{24}_L\rangle \cr\cr
& & +\langle \psi^{14 \ T}_L C \psi^{23}_L\rangle \Big ] \ . 
\label{a2fbar_su5condensate}
\eeqs
The fermions involved in this condensate gain dynamical masses of order
$\Lambda_5$, as do the nine gauge bosons in the coset SU(5)/SU(4). In addition
to breaking the SU(5) gauge symmetry, the condensate has the nonzero value of
the U(1)$'$ charge $Q'^{(2)} = -2$ given by the $k=2$ special case of
Eq. (\ref{q_akfbar_channel}) and hence breaks the global U(1)$'$ symmetry. Since
this symmetry is not gauged, this breaking yields one Nambu-Goldstone boson
(NGB).

To construct the low-energy effective field theory with SU(4) chiral gauge
invariance that describes the physics as the scale $\mu$ decreases below
$\Lambda_5$, we decompose the fermion representations of SU(5) with respect to
the unbroken SU(4) subgroup.  It will be useful to give this decomposition more
generally for SU($N$) relative to an SU($N-1$) subgroup in our usual notation
and also in terms of the corresponding Young tableaux:
\beqs
& & [2]_N = \{ [2]_{N-1} + [1]_{N-1} \}  \ , \ i.e., \cr\cr
& & 
\asym_{\ {\rm SU}(N)} = [ \ {\asym} + \fund \ ]_{\ {\rm SU}(N-1)} \ . 
\label{a2ndecomposition} 
\eeqs
The $[2]_4$ field is comprised of $\psi^{ab}_L$ fermions with 
$1 \le a, \ b \le 4$ 
that gained dynamical masses of order $\Lambda_5$ and were integrated out of
the low-energy theory.  The other massless SU(4)-nonsinglet fermions are the
$[1]_4 = F$ fermion $\psi^{5b}_L$ with $1 \le b \le 4$ and the
$\overline{[1]}_4 = \bar F$ fermion $\chi_{a,1,L}$ with $1 \le a \le 4$.
Hence, the massless SU(4)-nonsinglet fermion content of this theory consists of
$F + \bar F$, so this theory is vectorial. This SU(4) theory also contains the
SU(4)-singlet fermion $\chi_{5,1,L}$.  The one-loop and two-loop coefficients
of the SU(4) beta function have the same sign, so again, this function has no
IR zero, and therefore the SU(4) gauge coupling inherited from the SU(5) UV
theory continues to increase as the reference scale $\mu$ decreases.
Rewriting the left-handed $\bar F$ as a right-handed $F$, one sees that this is
a vectorial SU(4) gauge theory with massless $N_f=1$ Dirac fermion in the
fundamental representation.  It therefore has a classical global chiral 
flavor symmetry group ${\rm U}(1)_F \otimes {\rm U}(1)_{\bar F}$, or
equivalently, ${\rm U}(1)_V \otimes {\rm U}(1)_A$ in standard notation. The
U(1)$_A$ is broken by SU(4) instantons, so the nonanomalous global flavor
symmetry is ${\rm U}(1)_V$. At a scale $\Lambda_4 \lsim \Lambda_5$, one expects
that the SU(4) gauge interaction produces a bilinear fermion condensate in 
the most attractive channel, which is $F \times \bar F \to 1$, thus 
preserving the SU(4) gauge symmetry. The condensate is
\beq
\langle \sum_{b=1}^4 \psi^{5b \ T}_L C \chi_{b,1,L}\rangle \ . 
\label{su4_from_su5_condensate}
\eeq
This condensate respects the U(1)$_V$ global symmetry, and hence does not
produce any Nambu-Goldstone bosons.  Thus, this SU(4) theory confines and
produces gauge-singlet hadrons (with the baryons being bosonic).  In the
infrared limit, the only remaining massless particles are the SU(4)-singlet
fermion $\chi_{a,1,L}$ and the one Nambu-Goldstone boson resulting from the
breaking of the U(1)$'$ global flavor symmetry by the condensate 
(\ref{a2fbar_su5condensate}). 


\subsection{SU(6) $A_2 \, \bar F$ Theory}

We next consider an SU(6) $A_2 \, \bar F$ theory.  The fermion content of this
theory is the $N=6$ special case of (\ref{a2fbar}), namely 
$A_2+2\bar F = [2]_6 + 2\overline{[1]}_6$. The $A_2$ fermion is denoted 
$\psi^{ab}_L = -\psi^{ba}_L$,
and the two copies of the $\bar F$ fermion are denoted
$\chi_{a,i,L}$, where $1 \le a, \ b \le 6$ are gauge indices and $i=1,2$ is the
copy index.  We consider possible bilinear fermion condensates for this theory.
As discussed above, although the bilinear fermion condensation 
channels $A_2 \times A_2 \to \bar A_2$ and $A_2 \times \bar F \to F$ 
have the same $\Delta C_2$, a vacuum
alignment argument favors the $A_2 \times \bar F \to F$ channel because it
leaves a larger residual gauge symmetry, namely SU(5).  Assuming that a 
condensate in this channel does form, we denote the scale at which it is 
produced as $\Lambda_6$.  Again, by convention
we take the breaking direction as $a=6$ and the copy index as $i=2$ on the
$\bar F$ fermion in the condensate, which can thus be written as
\beq
\langle \sum_{b=1}^5 \psi^{6b \ T}_L C \chi_{b,2,L}\rangle \ . 
\label{a2fbar_condensate_N6}
\eeq
This condensate also breaks the 
${\rm SU}(2)_{\bar F} \otimes {\rm U}(1)'$ global flavor symmetry. 
The $\psi^{6b}_L$ and $\chi_{b,2,L}$ fermions with $1 \le b \le 5$ involved in
the condensate (\ref{a2fbar_condensate_N6}) get dynamical masses of order
$\Lambda_6$, as do the 11 gauge bosons in the coset SU(6)/SU(5).  These are
integrated out of the low-energy effective SU(5)-invariant theory that
describes the physics as the scale $\mu$ decreases below $\Lambda_6$. 

From the $N=6$ special case of the general decomposition
(\ref{a2ndecomposition}) in conjunction with the form of the condensate
(\ref{a2fbar_condensate_N6}), it follows that the massless SU(5)-nonsinglet
fermion content of the descendant SU(5) theory is $A_2+\bar F$, together with
the (massless) SU(5)-singlet fermions $\chi_{6,1,L}$ and $\chi_{6,2,L}$. Thus,
the SU(5)-nonsinglet fermion content of this theory is the same as that of the
SU(5) theory discussed above, and our analysis there applies here.  Since this
SU(5) theory satisfies the 't Hooft global anomaly matching conditions, when it
becomes strongly coupled, it could confine and produce massless SU(5)-singlet
composite fermions, as well as massive mesons and primarily gluonic states, or
it could self-break via fermion condensate formation.  We also discuss below a
possible SU(5)-preserving four-fermion condensate that might form.


\subsection{SU($N$) $A_2 \, \bar F$ Theories with $N \ge 7$ } 

For $N \ge 7$, the most attractive channel for bilinear fermion condensation 
is $A_2 \times \bar F \to F$, with $\Delta C_2$ 
given by the $k=2$ special case of (\ref{deltac2_akfbar_to_akminus1_channel}), 
\beq
\Delta C_2 = C_2([2]_N) = \frac{(N-2)(N+1)}{N} \quad {\rm for} \ 
A_2 \times \bar F \to F \ . 
\label{deltac2_a2fbar_channel}
\eeq
The UV to IR evolution of these theories is similar to that of
the SU(6) theory. At each stage, owing to the fact that the
SU($N$) theory and the various descendant theories satisfy 't Hooft
global anomaly matching conditions, as the coupling gets strong in the IR, 
the gauge interaction may confine and produce massless composite 
fermions or may produce various fermion condensates.  The most attractive
channel for bilinear fermion condensation at a given stage is 
$A_2 \times \bar F \to F$, breaking the theory down to the next descendant
low-energy theory.  If the theory follows the first type of UV to IR flow,
namely confinement with massless composite fermions, this extends all the way
to the IR limit, while if the theory follows the second type of flow with
condensate formation, then there is, in general, a resultant sequence of 
low-energy effective theories that describe the physics of the massless
dynamical degrees of freedom at lower scales. If all of the stages involve 
gauge (and global) symmetry breaking by fermion condensates, then the gauge 
symmetry breaking is of the form 
\beq
{\rm SU}(N) \ \to \ {\rm SU}(N-1) \to ... \to {\rm SU}(4) \ . 
\label{a2fbar_descent_sequence}
\eeq
Here, the last theory, namely the SU(4) theory, is vectorial, while all of the
higher-lying theories are chiral gauge theories.


\section{ $A_3 \, \bar F$ Theories}
\label{a3fbar_theories}

The fermion content of the $A_3 \, \bar F$ theories was displayed in
Eq. (\ref{a3fbar}). The one-loop and two-loop coefficients in the beta function
were given in Eqs. (\ref{b1_a3fbar}) and (\ref{b2_a3fbar}), with numerical
results for $\bar b_1$ and $\bar b_2$ displayed in Table
\ref{akfbar_properties}.  As is evident in Table \ref{akfbar_properties}, for
$7 \le N \le 10$, the coefficient $\bar b_2$ is positive, so the two-loop beta
function has no IR zero, and hence, as the reference scale $\mu$ decreases from
large values in the UV toward the IR, the gauge coupling increases until it
exceeds the region where it is perturbatively calculable. These theories are
thus strongly coupled in the infrared (marked as SC in Table 
\ref{akfbar_properties}). 

The next step in the analysis of the UV to IR flow in these theories is to
determine if one or more of them might satisfy the 't Hooft global anomaly
matching conditions.  If this were to be the case, then, as in the $A_2 \, \bar
F$ theories, one would have a two-fold possibility for the strongly coupled IR
physics, namely confinement with gauge-singlet composite fermions but no
spontaneous chiral symmetry breaking or formation of bilinear fermion
condensates with associated breaking of gauge and global symmetries.  For this
purpose, we have examined possible SU($N$) gauge-singlet fermionic operator
products to determine if any of them could satisfy these global anomaly
matching conditions.  The global flavor symmetry group was given in
Eq. (\ref{gflk3}) with (\ref{qveck3}).  We have not found any such fermionic
operator products. As an illustration of our analysis, let us consider the case
$N=7$, which contains a $[3]_7$ fermion $\psi^{abd}_L$ and two fermions,
$\chi_{a,i,L}$ with $i=1,2$ comprising two copies of the $\overline{[1]}_7$
representation. For this case,
\beq
G_{fl,N=7}^{(3)} = {\rm SU}(2)_{\bar F} \otimes {\rm U}(1)' \ , 
\label{gfl_k3n7}
\eeq
with ${\vec Q}' = (1,-5)$.    A fermionic operator product that is an 
SU(7) singlet is of the form 
\beq
f_{i,R} = \epsilon_{abdefgh}[\psi^{abd \ T}_L C \psi^{efg}_L] (\chi^c)^h_{i,R} 
\ , 
\label{a3fbar_fi}
\eeq
where the $c$ superscript denotes the charge conjugate fermion field. However,
this vanishes identically.  This can be seen as follows: an interchange
(transposition) of $\psi^{abd}_L$ and $\psi^{efg}_L$ entails a minus sign from
the switching of an odd number of indices in the antisymmetric SU(7) tensor
density, a second minus sign from Fermi statistics, and a third minus sign from
the fact that $C^T = -C$ for the Dirac charge conjugation matrix, so the
operator is equal to minus itself and hence is zero.  

Therefore, when theory becomes strongly coupled in the infrared, we will focus
on the type of UV to IR evolution that leads to fermion condensates, and we
consider bilinear fermion condensates here.  The most attractive channel for
these condensates, as a function of $N$, was given in Eq. (\ref{k3mac}).

As an explicit example of the $A_3 \, \bar F$ class of chiral gauge theories,
let us consider the SU(7) theory, which has chiral
fermion content given by the $N=7$ special case of Eq. (\ref{a3fbar}), namely
\beq
A_3 + 2\bar F = [3]_7 + 2\overline{[1]}_7 \ . 
\label{su7a3fbar}
\eeq
The most attractive channel for this theory is $A_3 \times A_3 \to \bar F$,
which breaks the gauge symmetry SU(7) to SU(6) and also breaks the global
flavor symmetry group ${\rm SU}(2)_{\bar F} \otimes {\rm U}(1)'$.  We denote
the scale at which this condensate forms as $\Lambda_7$. Without loss of
generality, we label the gauge index for the broken direction to be $a=7$. The
condensate then has the form
\beq
\langle \epsilon_{abdefg7} \psi^{abd \ T}_L C \psi^{efg}_L\rangle \ . 
\label{a3a3_to_fbar_condensate}
\eeq
Of the ${7 \choose 3}=35$ components of the $A_3$ fermion, denoted generically
as $\psi^{abd}_L$, the ${7 \choose 3}-{6 \choose 2} = 20$ components 
with $1 \le a, \ b, \ d \le 6$ that are involved in this
condensate gain dynamical masses of order $\Lambda_7$, as do the 13 gauge
bosons in the coset SU(7)/SU(6). These are integrated out of the low-energy
effective theory SU(6) chiral gauge theory that describes the physics as the
scale decreases below $\Lambda_7$.

The massless SU(6)-nonsinglet fermion content of this SU(6) theory thus
consists of $A_2 + 2 \bar F = [2]_6 + 2 \overline{[1]}_6$, comprised by the
${6 \choose 2}=15$ components $\psi^{ab7}_L$ and the $\chi_{a,i,L}$ with 
$1 \le a, \ b \le 6$ and $i=1,2$.  A theorem proved in \cite{cgtsab} states
that a low-energy effective theory that arises by dynamical symmetry breaking
from an (asymptotically free) anomaly-free chiral gauge theory is also
anomaly-free.  One sees that the present example is in accord with this general
theorem.  Indeed, the nonsinglet fermions in this SU(6) descendant theory are
precisely those of the SU(6) $A_2 \, \bar F$ theory discussed above, and that
analysis applies here for the further UV to IR evolution of the theory.  In
addition to the SU(6)-nonsinglet fermions, this descendant theory also contains
the SU(6)-singlet fermions $\chi_{7,i,L}$ with $i=1,2$.


\section{ $A_4 \, \bar F$ Theories}
\label{a4fbar_theories}

The fermion content of the $A_4 \, \bar F$ theories was given in
Eq. (\ref{a4fbar}).  The reduced one-loop and two-loop coefficients in the beta
function were listed in Eqs. (\ref{b1_a4fbar}) and (\ref{b2_a4fbar}), with
numerical results displayed in Table \ref{akfbar_properties}.  We find that for
each of the three relevant values of $N$, namely $N=9, \ 10, \ 11$, the
coefficient $\bar b_2$ is negative, so the two-loop beta function has an IR
zero.  As we noted above, for $N=11$, this IR zero is at very weak coupling
relative to the minimal critical value for bilinear fermion condensation, so we
can reliably conclude that the theory evolves from the UV to a (deconfined)
non-Abelian Coulomb phase in the infrared.  In the $N=9$ and $N=10$ theories,
the respective IR zeros in the two-loop beta function occur at strong and
moderate coupling, so a full analysis is necessary.

We have examined whether there are SU($N$) gauge-singlet composite fermion
operators that could satisfy the 't Hooft global anomaly matching conditions,
but we have not found any. The global flavor symmetry group was given in
Eq. (\ref{gflk4}) with (\ref{qveck4}).  As an illustration of our analysis, let
us consider the SU(9) $A_4 \, \bar F$ theory, which contains a $[3]_9$ 
fermion $\psi^{abd}_L$ and
the fermions, $\chi_{a,i,L}$ with $1 \le i \le 5$ comprising five copies of the
$\overline{[1]}_9$ representation of SU(9).  The global flavor symmetry group
is
\beq
G_{fl,N=9}^{(4)} = {\rm SU}(5)_{\bar F} \otimes {\rm U}(1)' \ , 
\label{gfl_k4n9}
\eeq
The $\chi_{a,i,L}$ fermions transform as $\overline{\fund}$ of the 
SU(5)$_{\bar F}$ flavor group, and the vector of U(1)$'$ charges is
${\vec Q}' = (Q'_{A_4},Q'_{\bar F}) = (1,-7)$. A fermionic operator product 
that is an SU(9) gauge singlet is 
\beq
f^i_R = \epsilon_{abdefghrs}[\psi^{abde \ T}_L C \psi^{fghr}_L] 
(\chi^c)^{s,i}_R \ .
\label{a4fbar_fi}
\eeq
This transforms as a $\fund$ representation of the global SU(5)$_{\bar F}$
symmetry with U(1)$'$ charge $2Q'_{A_4}-Q'_{\bar F} = 9$. Since this is a
right-handed composite fermion, we actually calculate with the charge conjugate
$(f^c)_{i,L}$, which is a left-handed fermion that transforms as a
$\overline{\fund}$ representation of the global SU(5) with U(1)$'$ charge $-9$.
We find that this composite fermion does not satisfy the global anomaly
matching conditions.  For example, consider the SU(5)$^3$ anomaly. The
fundamental fields make the following contributions: the $A_4$ fermion yields
zero, while the $\bar F$ fermions yield
$N {\cal A}(\overline{\fund})=9 \times (-1) = -9$.  However, the $f^c_L$ 
fermion yields ${\cal A}(\overline{\fund})=-1$, which does not match.  
Since we have not found
composite fermion operators that satisfy the 't Hooft global anomaly matching
conditions, we consider fermion condensation in the cases where the beta
function has an IR zero at moderate (for $N=10$) and strong (for $N=9$)
coupling in the infrared.

As an explicit example, we analyze the SU(9) $A_4 \, \bar F$ theory. The 
fermion content of this theory is given by the $N=9$ special case of 
Eq. (\ref{a4fbar}), namely
\beq
A_4 + 5\bar F = [4]_9 + 5\overline{[1]}_9 \ .
\label{su9ak4fbar}
\eeq
The most attractive channel for bilinear fermion condensation is the $N=9$
special case of (\ref{akak_to_fbar_channel}), namely 
$A_4 \times A_4 \to \bar F$.  Assuming that this condensate forms, it breaks 
the gauge symmetry SU(9) to SU(8) and also breaks the global flavor symmetry 
group ${\rm SU}(5)_{\bar F} \otimes {\rm U}(1)'$.  
We denote the scale at which this condensate forms as $\Lambda_9$. 
Without loss of generality, we label the gauge index for the
broken direction to be $a=9$. The condensate then has the form
\beq
\langle \epsilon_{abdefghr9} \psi^{abde \ T}_L C \psi^{fghr}_L\rangle \ . 
\label{a4a4_to_fbar_condensate}
\eeq
Of the ${9 \choose 4}=126$ components of $\psi^{abde}_L$, the 
${9 \choose 4} - { 8 \choose 3} = 70$ components
with $1 \le a, \ b, \ d, \ e \le 8$ that are involved in this
condensate gain dynamical masses of order $\Lambda_9$, as do the 17 gauge
bosons in the coset SU(9)/SU(8). These are integrated out of the low-energy
effective theory SU(8) chiral gauge theory that describes the physics as the
scale decreases below $\Lambda_9$.

The massless SU(8)-nonsinglet fermion content of this SU(8) theory thus
consists of $A_3 + 5 \bar F = [3]_8 + 5 \overline{[1]}_8$, comprised by the
${8 \choose 3}=56$ components $\psi^{abd9}_L$ and the $\chi_{a,i,L}$ with
$1 \le a, \ b, \ d \le 8$ and $1 \le i \le 5$.  Again, the theorem proved in 
\cite{cgtsab} guarantees that this SU(8) descendant theory is anomaly-free. 
Indeed, the nonsinglet fermions in this SU(8) descendant theory are
precisely those of the SU(8) $A_3 \, \bar F$ theory discussed above, and that
analysis applies here for the further UV to IR evolution of the theory.  In
addition to the SU(8)-nonsinglet fermions, this descendant theory also contains
the SU(6)-singlet fermions $\chi_{9,i,L}$ with $1 \le i \le 5$. 


\section{Multifermion Condensates and Implications for the Preservation of 
Chiral Gauge Symmetry} 
\label{multifermion_condensates}

Our discussion above of fermion condensate formation focused on bilinear
fermion condensates and resultant dynamical chiral gauge symmetry breaking.
However, it is, in principle, possible for a strongly interacting vectorial or
chiral gauge theory to produce fermion condensates involving product(s) of more
than just two fermion fields \cite{lacaze,su2an}.  Much less attention as been
devoted in the literature to such multifermion condensates than to bilinear
fermion condensates.  This is somewhat analogous to the situation with bound
states of (anti)quarks in hadronic physics.  For many years the main focus of
research was on color-singlet bound states with the minimum number of
(anti)quarks, namely $qqq$, for baryons and $q\bar q$ for
mesons. (Subsequently, glueballs and mixing between $q \bar q$ mesons and glue
to form mass eigenstates were also studied.) However, there is increasing
experimental evidence that the hadron spectrum also contains bound states with
additional quarks, such as $q \bar q q \bar q$ and $q \bar q Q \bar Q$, where
$Q$ means a heavy quark, $c$ or $b$, including charged mesons, and possibly
$qqq q \bar q$ and $qqq Q \bar Q$ \cite{multiquarkhadrons}.  In the case of
possible condensates involving four or more fermions, we are not aware of a
reliable method that can be used to assess the relative likelihood that these
would form. The problem of assessing this likelihood is fraught with even more
theoretical uncertainty than the uncertainty inherent in the use of the rough
MAC criterion to measure the attractiveness of bilinear fermion condensation
channels.

Clearly, Lorentz invariance implies that the number of
fermion fields in such multifermion condensates must be even.  As usual, we
denote the charge conjugate of a generic fermion field $\chi$ as $\chi^c \equiv
C \bar \chi^T$, where $C$ is the Dirac charge conjugation matrix satisfying 
$C = -C^T$ and $\bar \chi \equiv \chi^\dagger \gamma_0$; recall also that for a
left-handed fermion $\chi_L$, the charge conjugate is $(\chi_L)^c=(\chi^c)_R$. 

As an example, consider the SU(5) $A_2 \, \bar F$
theory, with the fields $\psi^{ab}_L$ and $\chi_{a,1,L}$ or equivalently, 
$\psi^c_{ab,R}$ and $(\chi^c)^{a,1}_R$. When the gauge 
interaction becomes strong, it could produce several different four-fermion
condensates that preserve the SU(5) gauge symmetry. One such condensate that
involves all of the fermions is 
\beq
\langle \epsilon_{abdef} [\psi^{ab \ T}_L C \psi^{de}_L] 
[\psi^{fs \ T}_L C \chi_{s,1,L}] \rangle \ ,
\label{su5fourfermioncondensate}
\eeq
where here $a,b,d,e,f,s$ are SU(5) gauge indices.  This condensate has 
U(1)$'$ charge $3Q'_{A_2} + Q'_{\bar F}$. Using the results from the $N=5$
special case of Eq. (\ref{qveck2}), namely, $Q'_{A_2}=1$, $Q'_{\bar F}=-3$, we
find that this condensate (\ref{su5fourfermioncondensate}) has zero
U(1)$'$ charge, so it also preserves the global U(1)$'$ symmetry of the SU(5)
theory.

In a similar manner, consider the SU(6) $A_2 \, \bar F$ theory, with the
fermions $\psi^{ab}_L$ and $\chi_{a,j,L}$ with $j=1,2$. As the
SU(6) gauge interaction becomes strong in the infrared, it might 
produce the following four-fermion condensate that is invariant under the SU(6)
gauge symmetry:
\beqs
& & \langle \epsilon_{abdesu}[\psi^{ab \ T}_L C \psi^{de}_L]
[(\chi^c)^{s,1 \ T}_R C (\chi^c)^{u,2}_R] \rangle - 
(1 \ \leftrightarrow \ 2) \cr\cr
& = & \langle  \epsilon_{abdesu}[\psi^{ab \ T}_L C \psi^{de}_L] 
\epsilon_{ij} [(\chi^c)^{s,i \ T}_R C (\chi^c)^{u,j}_R] \rangle \ .
\label{su6fourfermioncondensate}
\eeqs
Note that because of the contraction of the operator product 
$[(\chi^c)^{s,1 \ T}_R C (\chi^c)^{u,2}_R]$ with the SU(6) 
$\epsilon_{abdesu}$ tensor,
the first term in Eq. (\ref{su6fourfermioncondensate}) is automatically
antisymmetrized in the flavor indices $j=1,2$; we have made this explicit by
subtracting the term with these indices interchanged.  As shown by the second
line of Eq. (\ref{su6fourfermioncondensate}), this condensate thus preserves
the SU(2)$_{\bar F}$ factor group in the global flavor symmetry $G_{fl}$ for
this theory, namely ${\rm SU}(2)_{\bar F} \otimes {\rm U}(1)'$.  In the
$(A_2,\bar F)$ basis, the U(1)$'$ charges are $(2,-4)$, as given by the $N=6$
special case of Eq.  (\ref{qveck2}).  Hence, the U(1)$'$ charge of the
condensate (\ref{su6fourfermioncondensate}) is $-4$, so it breaks the U(1)$'$
part of $G_{fl}$, yielding one Nambu-Goldstone boson. 

One can give corresponding discussions of gauge-invariant multifermion
condensates for other SU($N$) $A_k \, \bar F$ theories that become strongly
coupled in the infrared.  In general, these theories could also 
produce other types of four-fermion condensates such as 
\beq
\langle [ \psi^{ab \ T}_L C \chi_{b,i,L} ]
        [(\psi^c)^T_{ad,R} C (\chi^c)^{d,j}_R]\rangle \ , 
\label{cfcfbar}
\eeq

\beq
\langle [\bar \psi_{ab,L} \gamma_\mu \psi^{ab}_L]
        [\bar \psi_{de,L} \gamma^\mu \psi^{de}_L ] \rangle \ , 
\label{psipsi}
\eeq
\beq
\langle [\bar \psi_{ab,L} \gamma_\mu \psi^{ab}_L]
        [(\bar \chi_L)^{d,i} \gamma^\mu \chi_{d,j,L}] \rangle \ , \
\label{psichi}
\eeq
and
\beq
\langle [(\bar \chi_L)^{a,i} \gamma_\mu \chi_{a,j,L}]
        [(\bar \chi_L)^{b,k} \gamma^\mu \chi_{b,\ell,L}] \rangle \ ,
\label{chichi}
\eeq
where $1 \le i, \ j, \ k, \ \ell \le n_{\bar F}$.  There are also multifermion
condensates with eight and more fermions that one could consider. 
Such multifermion condensates merit further study.


\section{Non-Existence of Asymptotically Free $S_k \, \bar F$ Theories with $k
  \ge 3$} 
\label{no_skfbar_kge3}

It is natural to carry out an investigation of (anomaly-free) chiral gauge
theories with gauge group SU($N$) and chiral fermions transforming according to
the rank-$k$ symmetric tensor representation with $k \ge 3$ and a requisite 
number of chiral fermions in the $\bar F$ representation so as to render the 
theories free of an anomaly in gauged currents.  We denote such a theory as an
SU($N$) $S_k \, \bar F$ theory.  This investigation would be the analogue of
the study that we have performed in this paper for $A_k \, \bar F$ theories
with $k \ge 3$ and would generalize the studies that have been carried out in
the past on the $S_2 \, \bar F$ theory \cite{by,ads,dfcgt,cgt,cgt2}.  As with
the $A_k \, \bar F$ theories, we require that the theory must be asymptotically
free so that it is perturbatively calculable in at least one regime, namely
the deep UV, where the gauge coupling is small. 

However, we shall show here that there are no asymptotically free
(anomaly-free) $S_k \, \bar F$ chiral gauge theories with $k \ge 3$.  As before
we denote the number of copies of $\bar F$ fermions as $n_{\bar F}$.  The
contribution to the triangle anomaly in gauged currents of a chiral fermion in
the $S_k$ representation is (see Appendix \ref{group_invariants} )
\beq
{\cal A}(S_k) = \frac{(N+k)! \, (N+2k)}{(N+2)! \, (k-1)!} \ . 
\label{anomsk}
\eeq
The total anomaly in the theory is
${\cal A} = {\cal A}(S_k)-n_{\bar F} {\cal A}(F)$, so 
the condition of anomaly cancellation is that 
\beq
n_{\bar F} = {\cal A}(S_k) \ . 
\label{nfbarsk}
\eeq
The first few values are of $n_{\bar F}$ are
\beq
n_{\bar F} = \cases{ N+4               & if $k=2$ \cr
                 (1/2)(N+3)(N+6)       & if $k=3$ \cr
                 (1/6)(N+3)(N+4)(N+8)  & if $k=4$ }
\label{nvalues_sk}
\eeq
and so forth for higher $k$. 

To investigate the restrictions due to the requirement of asymptotic freedom,
we calculate the one-loop coefficient of the beta function. We find 
\begin{widetext}
\beq
b_{1,S_k \bar F} = 
\frac{1}{3}\bigg [ 11N -2\Big \{ T(S_k) + {\cal A}(S_k) T(\bar F) \Big \} 
\bigg ] 
 = \frac{1}{3}\bigg [ 11N - \frac{1}{(k-1)!}\Big \{ 
\Big [\prod_{j=2}^k (N+j) \Big ]+\frac{(N+k)!(N+2k)}{(N+2)!} \Big \} \bigg ]
 \ . 
\label{b1akfbar_sk}
\eeq
\end{widetext}
We exhibit the explicit expressions for $b^{(k)}$ for the first few $k\ge 2$:
\beq
b_{1,S_2 \bar F} = 3N+2 \ , 
\label{b1_s2fbar}
\eeq
\beq
b_{1,S_3 \bar F} = \frac{1}{3}(-N^2+4N-12) \ , 
\label{b1_s3fbar}
\eeq
\beq
b_{1,S_4 \bar F} = -\frac{1}{9}(N^3+12N^2+14N+60) \ , 
\label{b1_s4fbar}
\eeq
\beqs
b_{1,S_5 \bar F} & & = -\frac{1}{36}(N^4+18N^3+119N^2+210N+360) \ , 
\cr\cr
& & 
\label{b1_s5fbar}
\eeqs
and
\beqs
b_{1,S_6 \bar F} & = & -\frac{1}{180}\Big (N^5+25N^4+245N^3+1175N^2 \cr\cr
                 & + & 2094N+2520 \Big ) \ . 
\label{b1_s6fbar}
\eeqs
The coefficient $b_{1,S_2 \bar F}$ is positive for all relevant $N$, and this
property was used in past studies of the $S_2 \, \bar F$ theory.  However,
the coefficient $b_{1,S_3 \bar F}$ is negative for relevant $N \ge 3$.  (Recall
that an SU(2) theory has only real representations and hence is not chiral.)
With $N$ generalized from positive integers to real numbers, $b_{1,S_3 \bar F}$
is negative for all $N$, reaching its maximum value of $-8/3$ for $N=2$.  We
find that the $b_{1,S_k \bar F}$ coefficients with $k \ge 4$ are
negative-definite for all positive $N$ (either real or integer). This is
evident from the illustrative explicit expressions that we have given for 
$4 \le k \le 6$.  This completes our proof that there are no asymptotically 
free, (anomaly-free) $S_k \, \bar F$ chiral gauge theories with $k \ge 3$. 


\section{Conclusions}
\label{conclusions}

In summary, in this paper we have constructed and studied asymptotically free
chiral gauge theories with an SU($N$) gauge group and chiral fermions
transforming according to the antisymmetric rank-$k$ representation, $A_k$,
with $k=2, \ 3, \ 4$, and, for each $k$ and $N$, the requisite number of
copies, $n_{\bar F}$, of fermions transforming according to the conjugate 
fundamental representation, $\bar F$, of this group to render the theory
anomaly-free. For a given $k$, to get a theory that is chiral and has 
$n_{\bar F} \ge 1$, we take $N \ge 2k+1$.  
We have extended previous studies of the $A_2 \, \bar F$ theories
with further analysis of fermion condensation channels and 
sequential symmetry breaking and have presented a number of new results on the 
$A_k \, \bar F$ theories with $k \ge 3$. 
The $A_2 \bar F$ theories form an infinite
family with $N \ge 5$, but we have shown that the $A_3 \, \bar F$ and 
$A_4 \, \bar F$ theories are only asymptotically free for $N$ in the 
respective ranges $7 \le N \le 17$ and $9 \le N \le 11$, and that there are 
no asymptotically free $A_k \, \bar F$ theories with $k \ge 5$.  
We have investigated the types of ultraviolet
to infrared evolution for these $A_k \, \bar F$ theories and have found that,
depending on $k$ and $N$, they may lead in the infrared to a
non-Abelian Coulomb phase, or may involve confinement with massless
gauge-singlet composite fermions, or bilinear fermion condensation with
dynamical gauge and global symmetry breaking.  In two cases, namely
$(k,N)=(2,6), \ (3,12)$, in each of which two bilinear fermion condensation
channels are equally attractive, so the MAC criterion does not prefer one over
the other, we have applied vacuum alignment arguments to infer which channel is
preferred. We have also discussed multifermion condensates.  Finally, we have
shown that there are no asymptotically free, anomaly-free SU($N$) $S_k \, \bar
F$ chiral gauge theories with $k \ge 3$, where $S_k$ denotes the rank-$k$
symmetric representation.


\begin{acknowledgments}
This research was partially supported by the NSF grant NSF-PHY-13-16617.
\end{acknowledgments}


\begin{appendix}

\section{Beta Function Coefficients and Relevant Group Invariants}
\label{beta_function}

For reference, we list the one-loop and two-loop coefficients
\cite{b1,b2} in the beta function (\ref{beta}) for a non-Abelian
chiral gauge theory with gauge group $G$ and a set of chiral fermions
comprised of $N_i$ fermions transforming according to the representations 
$R_i$: 
\beq
b_1 = \frac{1}{3}\Big [ 11 C_2(G) - 2 \sum_{R_i}N_i T(R_i) \Big ]
\label{b1gen}
\eeq
and
\beq
b_2=\frac{1}{3}\Big [ 34 C_2(G)^2 - 
2\sum_{R_i} N_i \{ 5C_2(G)+3C_2(R_i)\} T(R_i) \Big ] \ . 
\label{b2gen}
\eeq
%


\section{Relevant Group Invariants}
\label{group_invariants}

We list below the group invariants that we use for the relevant case 
$G={\rm SU}(N)$.  Recall that the order of the Lie group SU($N$) 
(i.e., the number of infinitestimal generators of the associated Lie algebra) 
is $o({\rm SU}(N))=N^2-1$, and the order of the symplectic group
${\rm Sp}(2N)$ is $o({\rm Sp}(2N))=N(2N+1)$.  The antisymmetric rank-$k$
representation of SU($N$) is denoted $A_k \equiv [k]_N$. Some group invariants
are
\beq
\sum_{i,j=1}^{{\rm dim}(R)} 
{\cal D}_R(T_a)_{ij} {\cal D}_R(T_b)_{ji}=T(R) \delta_{ab} 
\label{tr}
\eeq
and
\beq
\sum_{a=1}^{o(G)} \sum_{j=1}^{{\rm dim}(R)} 
{\cal D}_R(T_a)_{ij} {\cal D}_R(T_a)_{jk}=C_2(R)\delta_{ik} \ , 
\label{c2r}
\eeq
where $T_a$ are the generators of $G$, and ${\cal D}_R$ is the matrix
representation ({\it Darstellung}) of $R$.  

For the adjoint representation, $adj$, $C_2(adj) \equiv C_2(G)$, and for
$G={\rm SU}(N)$, $C_2(G)=N$.  For the rank-$k$ antisymmetric representation of
SU($N$), $A_k \equiv [k]_N$, 
\beq
C_2([k]_N) = \frac{k(N-k)(N+1)}{2N} 
\label{c2_asymk}
\eeq
and
\beq
T([k]_N) = \frac{1}{2}{N-2 \choose k-1} \ . 
\label{tr_asymk}
\eeq
Thus, for $k=1$, $T([1]_N)=T(F)=1/2$ and, for $k \ge 2$, 
\beq
T([k]_N) = \frac{\prod_{j=2}^k (N-j)}{2(k-1)!}  \ . 
\label{tr_asymkb}
\eeq
For the rank-$k$ symmetric representation, $S_k$, 
\beq
C_2(S_k) = \frac{k(N+k)(N-1)}{2N}
\label{c2_symk}
\eeq
and
\beq
T(S_k) = \frac{\prod_{j=2}^k (N+j)}{2(k-1)!} \ . 
\label{tr_symk}
\eeq

The anomaly produced by chiral fermions transforming according to the 
representation $R$ of a group $G$ is defined as 
\beq
{\rm Tr}_R(T_a,\{T_b,T_c\}) = {\cal A}(R)d_{abc}
\label{anomgen}
\eeq
where the $d_{abc}$ are the totally symmetric structure 
constants of the corresponding Lie algebra.  Thus, 
${\cal A}(\fund)=1$ for SU($N$). For the symmetric and 
antisymmetric rank-$k$ tensor representations of 
SU($N$), the anomaly is, respectively \cite{anomtk},
\beq
{\cal A}(S_k) = \frac{(N+k)! \, (N+2k)}{(N+2)! \, (k-1)!} \ .
\label{anomskapp}
\eeq
and, for $1 \le k \le N-1$, 
\beq
{\cal A}(A_k) = \frac{(N-3)!(N-2k)}{(N-k-1)! (k-1)!}  \ . 
\label{anom_asymk}
\eeq
(Note that $[N]_N$ is the singlet, so ${\cal A}([N]_N)=0$.) Hence, in 
particular, 
\beq
{\cal A}([2]_N) = N-4 \ , 
\label{anom_asym2}
\eeq
\beq
{\cal A}([3]_N) = \frac{(N-3)(N-6)}{2} \ ,
\label{anom_asym3}
\eeq
and
\beq
{\cal A}([4]_N) = \frac{(N-3)(N-4)(N-8)}{3!} \ . 
\label{anomaly_ak4}
\eeq
From Eq. (\ref{anom_asymk}), there follows the recursion relation 
\beq
{\cal A}([k]_N) + {\cal A}([k+1]_N) = {\cal A}([k+1]_{N+1}) \quad {\rm for} \ 
1 \le k \le N-1 \ . 
\label{anom_asymk_recursion}
\eeq

\end{appendix} 


\newpage


\begin{widetext}

\begin{table}
\caption{\footnotesize{Some properties of SU($N$) $A_k \, \bar F$ chiral gauge
    theories. The quantities listed are $k$, $N$, $n_{\bar F}$, $\bar b_1$,
    $\bar b_2$, and, for negative $\bar b_2$, $\alpha_{_{IR,2\ell}}=-\bar
    b_1/\bar b_2$, $\alpha_{cr}$ for the most attractive bilinear fermion
    condensation channel (\ref{channel}) in the SU($N$) theory, and the ratio
    $\rho_c$. The dash notation $-$ means that the two-loop beta function has
    no IR zero.  The likely IR behavior is indicated in the last column, with
    the abbreviations SC, MC, WC for the type coupling in the IR (SC = strong,
    MC = moderate, WC = weak coupling). In the WC case, the UV to IR evolution
    is to a non-Abelian Coulomb phase (NACP). The various possibilities for the
    evolution involving strong and moderately strong coupling are discussed in
    the text.  For $k=2$, we include illustrative results covering the interval
    $5 \le N \le 10$; for $k=3, \ 4$ we list results for all (asymptotically
    free) $A_k \, \bar F$ theories.}}
\begin{center}
\begin{tabular}{|c|c|c|c|c|c|c|c|c|}
\hline\hline
$k$ & $N$  & $n_{\bar F}$ & $\bar b_1$ & $\bar b_2$ & $\alpha_{_{IR,2\ell}}$ 
& $\alpha_{cr}$ & $\rho_c$ & IR coupling 
\\ \hline
2 & 5 & 1  & 1.3528 & 1.4996  & $-$  & 0.44 & $-$ & SC \\
2 & 6 & 2  & 1.59155& 2.0486  & $-$  & 0.45 & $-$ & SC \\
2 & 7 & 3  & 1.8303 & 2.6796  & $-$  & 0.37 & $-$ & SC \\
2 & 8 & 4  & 2.0690 & 3.3927  & $-$  & 0.31 & $-$ & SC \\
2 & 9 & 5  & 2.3077 & 4.1879  & $-$  & 0.27 & $-$ & SC \\
2 &10 & 6  & 2.5465 & 5.0654  & $-$  & 0.24 & $-$ & SC \\
\hline
3 & 7  & 2 & 1.7242  & 2.1525    & $-$    & 0.20 & $-$ & SC \\
3 & 8  & 5 & 1.8038  & 1.9784    & $-$    & 0.21 & $-$ & SC \\
3 & 9  & 9 & 1.8303  & 1.3805    & $-$    & 0.21 & $-$ & SC \\
3 & 10 &14 & 1.8038  & 0.2573    & $-$    & 0.21 & $-$ & SC \\
3 & 11 &20 & 1.7242  & $-1.4926$ & 1.155  & 0.21 & 5.4 & SC \\
3 & 12 &27 & 1.59155 & $-3.9705$ & 0.4008 & 0.21 & 1.9 & MC \\
3 & 13 &35 & 1.4059  & $-7.2779$ & 0.1932 & 0.19 &0.99 & MC \\
3 & 14 &44 & 1.1671  & $-11.5161$& 0.1013 & 0.18 &0.57 & MC \\
3 & 15 &54 & 0.8753  & $-16.7864$& 0.05215& 0.16 & 0.32 & WC, NACP \\
3 & 16 &65 & 0.5305  & $-23.1901$& 0.02288& 0.15 & 0.15 & WC, NACP \\
3 & 17 &77 & 0.1326  & $-30.8287$& 0.00430& 0.14 & 0.03 & WC, NACP \\
\hline
4 & 9  & 5 & 1.5650  & $-0.5896$ & 2.6542 & 0.12 & 22.5 & SC  \\
4 & 10 &14 & 1.0610  & $-5.3310$ & 0.1990 & 0.12 & 1.7  & MC  \\
4 & 11 &28 & 0.2387  & $-13.410$ & 0.0178 & 0.12 & 0.15 & WC, NACP \\
\hline\hline
\end{tabular}
\end{center}
\label{akfbar_properties}
\end{table}
%


\begin{table}
\caption{\footnotesize{$\Delta C_2$ values for the SU($N$) 
    $A_k \, \bar F$ chiral
    gauge theories and most attractive channels for bilinear fermion
    condensation.  The quantities listed are $k$, $N$, and 
    the respective $\Delta C_2$ values for the 
    $A_k \times A_k \to \bar A_{N-2k}$ and 
    $A_k \times \bar F \to A_{k-1}$ channels. In the last column, we list the
    most attractive channel for bilinear fermion condensation in the strongly
    coupled and moderately strongly coupled (SC,MC) cases.  
    If the UV to IR evolution remains
    weakly coupled (WC), it flows to a non-Abelian Coulomb phase (NACP).
    For $k=2$, we include illustrative results including the interval 
    $5 \le N \le 10$; for $k=3, \ 4$ we list results for all (asymptotically
    free) $A_k \, \bar F$ theories. See text for further discussion of
    the $k=2$, $N=6$ and $k=3$, $N=12$ cases where the 
    $\Delta C_2$ values are equal. The  $A_2 \, \bar F$ theories could 
    confine, yielding massless composite fermions.  
    Possible multifermion condensates are also discussed in the text.}}
\begin{center}
\begin{tabular}{|c|c|c|c|c|}
\hline\hline
$k$ & $N$ & $\Delta C2(A_k \times A_k \to \bar A_{N-2k})$ & 
$\Delta C_2(A_k \times \bar F \to A_{k-1})$ &  MAC for (S,M)C \\ \hline
2 & 5  & 4.800    & 3.600   & $A_2 \times A_2 \to \bar F$ \\
2 & 6  & 4.667    & 4.667   & $A_2 \times \bar F \to F$  \\
2 & 7  & 4.571    & 5.714   & $A_2 \times \bar F \to F$ \\
2 & 8  & 4.500    & 6.750   & $A_2 \times \bar F \to F$ \\
2 & 9  & 4.444    & 7.778   & $A_2 \times \bar F \to F$ \\
2 &10  & 4.400    & 8.800   & $A_2 \times \bar F \to F$ \\
\hline
3 & 7  & 10.29    & 4.571   & $A_3 \times A_3 \to \bar F$ \\
3 & 8  & 10.125   & 5.625   & $A_3 \times A_3 \to \bar A_2$ \\
3 & 9  & 10.000   & 6.667   & $A_3 \times A_3 \to \bar A_3$ \\
3 & 10 & 9.900    & 7.700   & $A_3 \times A_3 \to \bar A_4$ \\
3 & 11 & 9.818    & 8.727   & $A_3 \times A_3 \to \bar A_5$ \\
3 & 12 & 9.750    & 9.750   & $A_3 \times \bar F \to A_2$ \ or \ NACP \\
3 & 13 & 9.692    & 10.769  & $A_3 \times \bar F \to A_2$ \ or \ NACP \\
3 & 14 & 9.643    & 11.786  & $A_3 \times \bar F \to A_2$ \ or \ NACP \\
3 & 15 & 9.600    & 12.800  & NACP \\
3 & 16 & 9.5625   & 13.8125 & NACP \\
3 & 17 & 9.529    & 14.824  & NACP \\
\hline
4 & 9  & 17.78    & 5.556   & $A_4 \times A_4 \to \bar F$ \\
4 & 10 & 17.60    & 6.600   & $A_4 \times A_4 \to \bar A_2$ \ or \ NACP \\
4 & 11 & 17.45    & 7.636   & NACP \\
\hline\hline
\end{tabular}
\end{center}
\label{kN_DeltaC2_values}
\end{table}

\end{widetext}


\begin{thebibliography}{99}
\bibitem{conjugates}
%
When no confusion would result about the value of $N$,
we often use the notation $A_k$ and $\bar F$ for $[k]_N$ and
$\overline{[1]}_N$. Our study of $A_k \, \bar F$ chiral gauge theories is
obviously equivalent to the study of the set with fermions in the conjugate
representations $\bar A_k$ and $F$.
\end{thebibliography}
\end{document}